\documentclass[aps,jcp,amsmath,amssymb,reprint]{revtex4-2}

\usepackage{chemformula} 
\usepackage[T1]{fontenc} 
\usepackage{multirow}
\usepackage{dcolumn}
\usepackage{tabularx}
\usepackage{float}
\usepackage{xr}
\usepackage{hyperref}
\usepackage[utf8]{inputenc}
\usepackage{mciteplus}
\usepackage{subfig}
\usepackage[utf8]{inputenc}
\setcitestyle{super}
\usepackage[cdot,mediumqspace,amssymb]{SIunits}

\usepackage{textcomp} 
\newcommand{\textapproxtilde}{\raisebox{0.5ex}{\texttildelow}} 

\newcommand{\comment}[1]{}

\newcommand*{\citen}[1]{%
  \begingroup
    \romannumeral-`\x 
    \setcitestyle{numbers}%
    \cite{#1}%
  \endgroup   
}

\begin{document}
	\preprint{AIP/123-QED}
	
	\title{One-electron self-interaction error and its relationship to geometry and higher orbital occupation}
	
	\author{Dale R. Lonsdale}
	\author{Lars Goerigk}%
	\email{lars.goerigk@unimelb.edu.au}
	\affiliation{School of Chemistry, The University of Melbourne, Victoria 3010, Australia}

\begin{abstract}

Density Functional Theory (DFT) sees prominent use in computational chemistry and physics, however, problems due to the self-interaction error (SIE) pose additional challenges to obtaining qualitatively correct results. An unphysical energy an electron exerts on itself, the SIE impacts most practical DFT calculations. We conduct an in-depth analysis of the one-electron SIE in which we replicate delocalization effects for simple geometries. We present a simple visualization of such effects, which may help in future qualitative analysis of the one-electron SIE. By increasing the number of nuclei in a linear arrangement, the SIE increases dramatically. We also show how molecular shape impacts the SIE. Two and three dimensional shapes show an even greater SIE stemming mainly from the exchange functional with some error compensation from the one-electron error, which we previously defined [{\it Phys. Chem. Chem. Phys.} {\bf 22}, 15805 (2020)]. Most tested geometries are affected by the functional error, while some suffer from the density error. For the latter we establish a potential connection with electrons being unequally delocalized by the DFT methods. We also show how the SIE increases if electrons occupy higher-lying atomic orbitals; seemingly one-electron SIE free methods in a ground are no longer SIE free in excited states, which is an important insight for some popular, non-empirical DFAs. We conclude that the erratic behavior of the SIE in even the simplest geometries shows that robust density functional approximations are needed. Our test systems can be used as a future benchmark or contribute towards DFT development.

\end{abstract}
\maketitle


\section{Introduction}
\label{sec:intro}
It is commonplace to accompany or predict molecular or solid-state chemistry and physics with quantum-chemical calculations. However, with the relative ease anyone can conduct such calculations, alongside continuing development in the field, we must be mindful of the capabilities and inadequacies of the methodologies we use. 

In the realm of molecular quantum chemistry we can broadly designate two camps in which many popular computational methods lie: Wave Function Theory (WFT) and Density Functional Theory (DFT).\cite{Hohenberg1964,KSDFT_orig} 
Shortcomings of the former are mostly a question of computational feasibility: including just enough electron correlation to capture the essential physics governing the system in question while getting the result in a timely fashion.
In contrast, DFT is generally computationally less intensive, but retains drawbacks which differ depending on which of the applicable forms of DFT is under consideration. 
Wildly popular is the Kohn-Sham DFT formalism\cite{KSDFT_orig} which sacrifices computational efficiency for the reintroduction of orbitals, thus yielding a level of guaranteed accuracy for the electronic kinetic energy. 
Instead, the primary shortcoming of DFT is finding accurate functional forms for the exchange and correlation terms, which still remain elusive. 
In lieu of an exact solution to the prior two terms, it becomes pragmatically justified to use density functional approximations (DFAs) --- all of which introduce errors.
It is the improper calculation of exchange which leads to one of the most pervasive pitfalls of DFT: the notorious self-interaction error (SIE).

For better or worse, routine computational chemistry applications use DFT approaches --- most of them suffering from varying degrees of SIE,\cite{Goerigk2017,2006_many_elecsie} often unbeknownst to mere DFT users.
This error subtly, though sometimes dramatically, changes the quantitative result of every practical DFT application and proves to be a barrier to computational modelling. 
Past attempts to separate and correct the SIE have had varying success,\cite{PZ_SIC_short_bonds} which is why one important aspect is to understand its empirical nature so that the users and developers of DFT methods can predict when their calculations have been influenced by this spurious energy contribution.\cite{2006_many_elecsie,Zhang1998,Lundberg2005,ErinSIERev}

Dedicating research efforts to the SIE is justified due to its ability to cause qualitative errors. For example in the case of the simplest hydrogen transfer reaction, \ch{H2 + H ->H + H2}, many DFAs such as BLYP\cite{BLYP_1_B88Ex,BLYP_2,BLYP_3} drastically underestimate this transition state energy,\cite{H3_barrier_underestimate} sometimes predicting other similar hydrogen abstraction reactions to be barrier-less processes.\cite{barrierless_paper} In such cases, finding the geometry of the transition state is fruitless due to the SIE's effect on the potential energy surface. Spurious proton transfers were reported for organic acid-base co-crystals due to the SIE/delocalization problem.\cite{LeBlanc2018} Furthermore, there are issues relating the SIE to applications including, but not limited to, charge transfer,\cite{CTproblem1,CTproblem2,CT_Dreuw_MHG,wB2PLYPwB2GPPLYP,CommentOnAdamoCT} magnetic properties,\cite{Magnetic_SIE} spin-state splittings/energetics,\cite{Swart_SpinStates1,Swart_SpinStates2,SSB-D,Coskun2016,Kao2017} transition state energies,\cite{TS_SIE_early_mid_late} thermochemistry,\cite{Johnson2008,goerigk2010,Goerigk2011,Goerigk2017} halogen\cite{OterodelaRoza2014,Burke_2018} and chalcogen bonding,\cite{Burke_2018,CHAL336} halogen-bond dissocation,\cite{Johnson2003} solvated electrons,\cite{SIEsolvatedElectrons} and dissociation of neutral species resulting in unphysically delocalized electrons.\cite{spurious_FracCharge2006,dissociation_diatomic_molecules}

One phenomenological description of the SIE's manifestation is that it is an artificial delocalization of a single electron over two or more nuclei at impossibly large distances.\cite{ErinSIERev,SIEsolvatedElectrons,spurious_FracCharge2006,Gilson2011,Johansson2008,Dutoi2006} The electron does not properly localize on one center and can instead have its negative charge stabilized over a longer distance. More formally, the SIE comes from the expression for the classical Coulomb integral:

\begin{equation}
J[\rho] = \frac{1}{2} \iint \frac{\rho(\vec{r}_1)\rho(\vec{r}_2)}{\vec{r}_{12}} \ d\vec{r}_1d\vec{r}_2 \ ,
\label{coulomb_energy}
\end{equation}

where $\rho$ is the electron density, $\vec{r}$ represents the spatial coordinates of an electron, and $\vec{r}_{12}$ is the distance between two electrons. 
For a single electron the Coulomb operator yields a non-zero value and leads to a repulsive-self interaction energy between one electron and itself.
This term also exists in WFT, but is cancelled out at the Hartree-Fock (HF) level by the corresponding exchange-based self-interaction integrals. As exchange is approximated in DFT, only an imperfect cancellation occurs, leading to the SIE.

For as long as approximate exchange functionals have been used, the SIE has been an issue which many have sought to solve; for a recent review, see Ref. \citen{ErinSIERev}. One of the most used explicit self-interaction corrections (SICs) comes from the orbital-by-orbital removal approach developed by Perdew and Zunger.\cite{perdew1981self} This approach has limited success due to shortcomings associated with the technique,\cite{SIC_BAD_thermo,PZ_SIC_short_bonds,SIC_is_goodBad} though other avenues to alleviate these issues are an ongoing source of research: some examples include scaling\cite{Vydrov2006,Bhattarai2021} and use of Fermi-\cite{FERMI_SIC,FLOSIC_pastLDA} or complex-orbitals.\cite{Hannes_First_cmplxSIC,HannesLehtolaMHG_cmplxSIC,HannesLehtola_effect_cmplxSIC} In the solid-state realm, the DFT+U correction\cite{1991dft+u_vladimir} is quite popular. However, simply correcting or removing the SIE is not necessarily a straightforward improvement; SIE often mimics important, non-dynamic electron correlation effects absent in conventional DFT calculations.\cite{poloElectronCorrelationSelfinteraction2002,hendersonConnectionSelfinteractionStatic2010} Therefore, one of the more straightforward solutions is the admixture of exact Fock-exchange in a class of DFAs known as hybrids.\cite{Becke1993} DFA exchange components can come from the Local Density Approximation (LDA), Generalized-Gradient Approximation (GGA), or meta-GGA (mGGA) forms. Hybrids are commonly employed as a kind of self-interaction correction, but using too much exact exchange inherits the problems associated with HF theory.

DFA development through benchmarking can create statistically improvable functionals that perform well across a range of test systems covering electronic ground\cite{Goerigk2011,Peverati2014,Goerigk2017,Mardirossian2017,Goerigk2019} and excited states.\cite{Silva-Junior2008a,Goerigk2009,Goerigk_2010,Leang2012,RevLaurent2013,wB2PLYPwB2GPPLYP,TDHDFAsTriplets,SCSRSDHDFAs,TDDHDFaccount}
Naturally, to truly fix the SIE and related issues we must also have advancement in satisfying constraints from first principles: see, Refs \citen{Kronik2020} and \citen{Kraisler2020} for examples. 
Fixing DFAs from a more fundamental level aims to solve problems like the SIE in an elegant manner --- perhaps leading to the next generation of electronic structure methods.
In lieu of this, we can attempt to dexterously avoid and identify the SIE through studies that characterize its manifestations.\cite{Zhang1998,Lundberg2005,spurious_FracCharge2006}
One such important distinction comes from the separation of the SIE into many- and one-electron components,\cite{2006_many_elecsie} of which, we will be only concerning ourselves with the one-electron part due to the difficulty in writing down a clear expression for the many-electron portion.

In fact, we already extensively analyzed the one-electron SIE in a previous paper\cite{Lonsdale2020} in which we used mono- and dinuclear one-electron systems to probe the various components on 74 different DFAs. 
We also thoroughly characterized the basis set dependence of the SIE and correlated it with the exponents of the applied Gaussian-type orbitals, which impact the latters' diffuseness.
We also demonstrated that the one-electron SIE is a sizeable component in several of the most accurate DFAs to date, and that there is an approximately linear relationship between the one-electron SIE and the nuclear charge. However, our previous, relatively simple model systems might be insufficient to generalize our findings for more realistic calculations. Therefore, we have set about extending our information on one-electron SIE to more complicated model systems, thus, hopefully extending these insights to real-world applications.

A critical task in computational chemistry is surveying potential energy surfaces for global minima and transition states prior to modelling chemical or physical properties. Just as the SIE can impact desired quantities directly, e.g. overly red-shifted excitation energies and spectra---see Refs \citen{RevLaurent2013} and\citen{TDDHDFaccount} for reviews on DFT for excited states --- it can also affect them indirectly through artificially stable or unstable geometries, e.g. failure to predict a transition state. As the structural arrangement of atoms often underpins fundamental chemical and physical properties of molecules and solids, sometimes even subtle alterations of predicted geometries can cause speciously-calculated properties. Through shifting the topology of the potential energy surface, the SIE can cause qualitative failures in the prediction of structures. This is why robustness in particular is such a highly prized feature of a DFA in many benchmark studies.\cite{Goerigk2017,SCSRSDHDFAs} Therefore, we find it valuable to consider the one-electron SIE in the context of different geometries to provide more boundaries around when our current methods could fail.

Additionally, how we have chosen to construct our one-electron SIE datasets in our previous study has left us with electronic structures corresponding to only the 1s atomic orbital or MOs formed from 1s orbitals.\cite{Lonsdale2020} As valence orbitals are usually the most important for describing chemical bonding and changes of the electronic structure during a reaction, it follows that the SIE picture will change in more realistic systems containing nuclei beyond helium.
Therefore, we have also sought to account for the impact the SIE has on higher-lying orbitals. In this way, we hope to incrementally build from the one-electron picture, a more realistic description of the SIE. 

Herein, we present results and discussion into the geometry aspect of the SIE, going through each system and the breakdown of errors within. Then, we visualise the SIE with density-difference plots and touch on SIE effects when increasing the nuclear charge in our various geometries. Finally, we connect the magnitude of the one-electron SIE to higher-lying 2s and 2p atomic orbitals. These new findings will provide further insights and data that can be used in the development of one-electron SIE corrections or SIE-free DFAs.

\subsection{Theoretical background}\label{sec:compdet}

We present only a brief series of definitions of the SIE and the breakdown into its components: a more detailed discussion can be found in the theoretical details of our previous study.\cite{Lonsdale2020}

As we consider only one-electron systems within the Born-Oppenheimer Approximation without relativistic effects, HF Theory correctly gives the exact energy and density for such systems. To account for the self-interaction energy from eq. \eqref{coulomb_energy}, a set of exchange integrals generate an energy equivalent in magnitude but opposite in sign --- the result is that these two quantities vanish and the exact-energy expression for one-electron systems becomes:

\begin{equation}\label{exactresult}
E^{HF}_{1el}[\rho]=h[\rho]\ ,
\end{equation}

where $h$ is the expectation value of the one-electron Hamiltonian --- stemming from electronic kinetic and electron-nucleus interaction energies --- and $\rho$ is the exact one-electron density. As many DFAs are not one-electron SIE free and only yield an approximate density and total energy we can express the SIE of such a functional as:

\begin{equation}\label{totalSIE2}
\begin{split}
SIE[\tilde\rho,\rho]&=E^{DFA}[\tilde\rho]-E^{HF}_{1el}[\rho]\\
      &=E^{DFA}[\tilde\rho]-h[\rho] \\
      &= J[\tilde\rho] +E_{X}^{DFA}[\tilde\rho]+E_{C}^{DFA}[\tilde\rho] + h[\tilde\rho] - h[\rho]\ ,
\end{split}
\end{equation}

where $\tilde\rho$ is the approximate electron density for the DFA, $J$ is the classic Coulomb repulsion energy, and $E_{X}^{DFA}$ and $E_{C}^{DFA}$ are the DFAs' exchange and correlation energies, respectively. 
We can then break the one-electron SIE down into various component parts which we briefly detail below.  Firstly, improper cancellation between the approximate DFA exchange and the Coulomb energy leads to the exchange SIE:

\begin{equation}\label{eqn.Ex}
E_X^{SIE}[\tilde\rho] = E_X^{DFA}[\tilde\rho] +J[\tilde\rho] \ .
\end{equation}
Note that we have dropped the subscript ${1el}$, as it is clear that the context of our SIE discussion is the one-electron case.
As electron correlation for one electron systems should be zero by definition, we can simply write the correlation energy as the correlation SIE:

\begin{equation}\label{eqn.Cor}
E_C^{SIE}[\tilde\rho] = E_C^{DFA}[\tilde\rho] \ .
\end{equation}

As $E_X^{DFA}$ and $E_C^{DFA}$ are approximate, applying them during an SCF procedure causes an incorrect density, and in turn an error from the formally exact one-electron term $h$, which we defined as the one-electron error (OEE):\cite{Lonsdale2020} 

\begin{equation}
OEE[\tilde\rho,\rho] = h[\tilde{\rho}] - h[\rho] \ .
\label{sie_OEE_err}
\end{equation}
The sum of eqs \ref{eqn.Ex}, \ref{eqn.Cor}, \& \ref{sie_OEE_err} returns the total SIE in eq. \ref{totalSIE2}.

 Not specific to the one-electron SIE, Sim, Burke, and co-workers\cite{Sim2014,Burke_2018_2} have broken down errors from DFAs into density- and functional-based components. This is an attempt to separate the error stemming from an incorrectly calculated density, through the self-consistent field (SCF) procedure, from the error that comes from a single non-SCF DFA calculation --- presupposing an accurately/exactly calculated density. We expanded upon this concept in our last paper \cite{Lonsdale2020} to specifically cover the one-electron SIE, so the functional error becomes: 
\begin{equation}
SIE^{func}[\rho] = E^{DFA}[\rho] - E^{HF}[\rho]
\label{sie_func_err}
\end{equation}
and the density-error is:
\begin{equation}
SIE^{dens}[\tilde\rho,\rho] = E^{DFA}[\tilde{\rho}] - E^{DFA}[\rho] \ .
\label{sie_dens_err}
\end{equation}

We would like to note these formulations of $SIE^{func}$ and $SIE^{dens}$ are only relevant in the context of the one-electron SIE and that they would need to be adapted to apply to many electron systems; see Ref \citen{Sim2022} for a recent appraisal of density-corrected DFT. The sum of eqs \ref{sie_func_err} and \ref{sie_dens_err} return the total one-electron SIE as defined in eq. \ref{totalSIE2}. The OEE from eq. \ref{sie_OEE_err} is, thus, related to and a part of the $SIE^{dens}$ from eq. \ref{sie_dens_err}.\cite{Lonsdale2020}

\subsection{Computational details}\label{subsec:computational}

Calculations were performed with QCHEM V5.3.1\cite{QchemRef}, V5.2 in the case of Sections \ref{Sec:HighZ} and \ref{sec:higherangle}, using unrestricted Kohn-Sham DFT and HF algorithms with the decontracted aug-cc-pVQZ basis set.\cite{Dunning1989,Kendall1992} This approach is the same as our previous study, so we refrain from a basis-set dependence study: see Ref \citen{Lonsdale2020} for details. 
In order to calculate some breakdowns of the SIE (see eqs \ref{sie_func_err} and \ref{sie_dens_err}) some of the DFT and HF calculations were followed by non-SCF procedures, where appropriate. 
For numerical quadrature we used an unpruned grid of 99 radial spheres and 590 angular spheres with standard convergence thresholds.
There were several cases of poor convergence, whereby we used standard convergence techniques as well as stability analysis, level shifting, and various damping variants. 

\begin{figure}
\begin{center}
\includegraphics[width=1\linewidth]{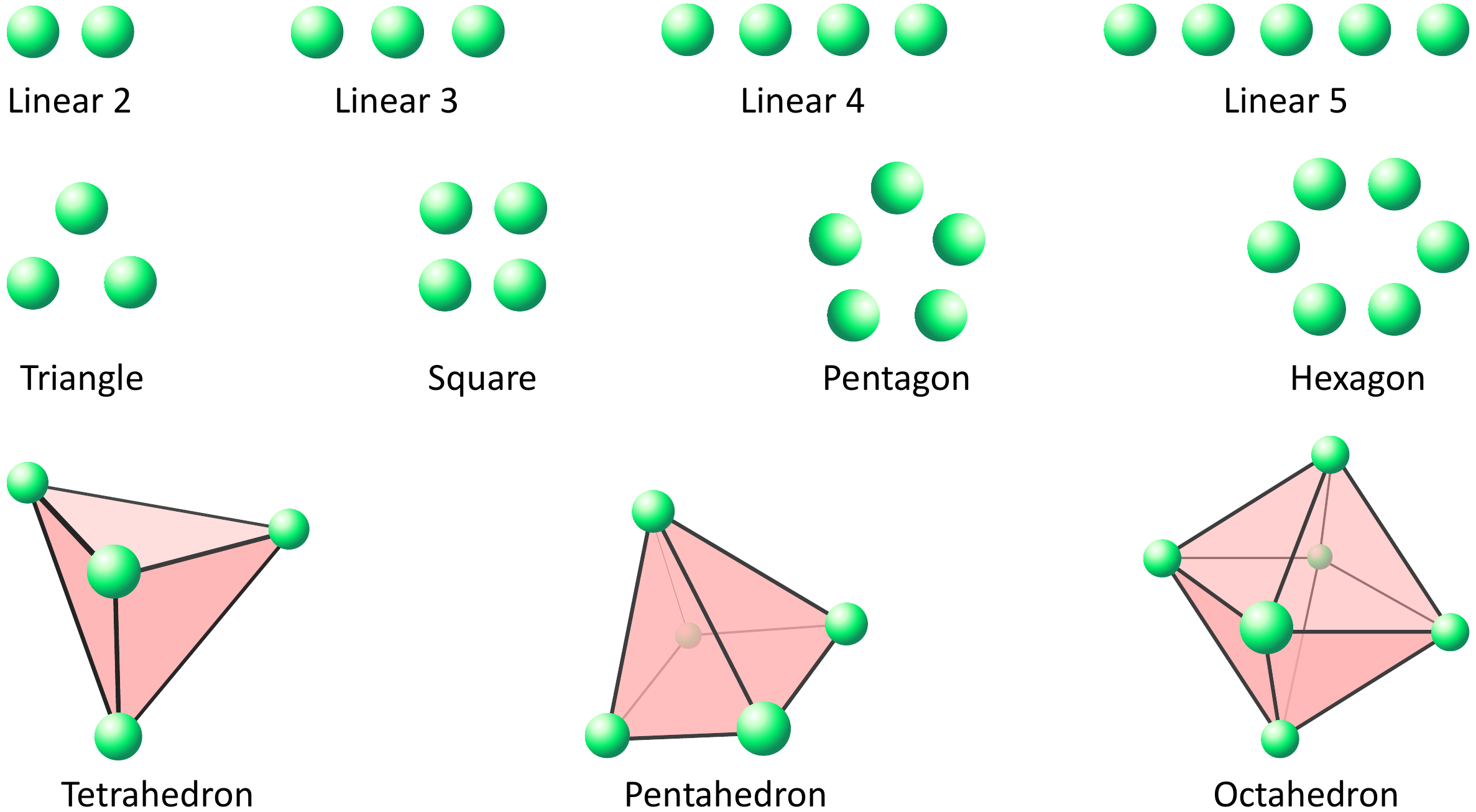}
\end{center}
\caption{Geometries investigated in this study. Polygons and polyhedra are regular; linear systems possess D$_{\infty \mathrm{h}}$ symmetry.}
\label{fig:geometries}
\end{figure}

To investigate the SIE's effect on a range of geometries we chose several common nuclear arrangements presented in Fig. \ref{fig:geometries}. We have split them into 3 categories: one-dimensional strings of 2, 3, 4, or 5 nuclei, dubbed ``\textbf{Linear 1}'', `\textbf{Linear 2}'' etc.;  two-dimensional shapes including a \textbf{Triangle}, \textbf{Square}, \textbf{Pentagon,} and \textbf{Hexagon}; and three-dimensional \textbf{Tetrahedron}, \textbf{Pentahedron}, and \textbf{Octahedron} geometries, again labelled in bold in the following discussion. These shapes were chosen as model systems because they can be regular, highly symmetric arrangements with each nucleus being equidistant to all its nearest neighbors. The equilateral side length in all polygons and polyhedra is denoted here as ``$r$''. For the strings of nuclei all internuclear distances between nearest neighbors were the same and also denoted as ``$r$''. By introducing such symmetry we can characterize the delocalized nature of the one-electron SIE through an idealized set of parameters. In order to study the SIE exactly, all systems possess exactly one electron. All atomic positions are predominately hydrogen nuclei --- except for Section (\ref{Sec:HighZ}) which considers the effect of higher nuclear charge. The $r$ value was increased to create dissociation curves for all the structures from 0.1 to 10 \angstrom\ in 100 points with equal spacing. 
In particular, we only calculated these points for the hydrogen nuclei, for nuclei helium to carbon we calculated $r$ values from 0.5 to 10 \angstrom\ in 20 equally spaced points. 

Due to the unusual nature of our test systems there were severe convergence issues for several DFAs we otherwise might have included.  That being said, the purpose of our study was not to rank a large number of DFAs, like we have done in our previous work, but to exemplify SIE-related behavior for geometries and higher-lying occupied orbitals. For that reason, we limited our selection to several common and/or unique functionals that had been insightful in our previous study.\cite{Lonsdale2020} DFAs tested on the various geometries were: SVWN5,\cite{xalpha,VWN} BLYP,\cite{BLYP_3,BLYP_2,BLYP_1_B88Ex} PBE,\cite{PBE} TPSS,\cite{TPSS} B97M-V,\cite{B97M-V} B3LYP,\cite{B3LYP_1,b3lypb} PBE0,\cite{pbe0_1,pbe0_2} SCAN0,\cite{SCAN0} BHLYP,\cite{Becke1993} B75LYP (75 \% Fock exchange), and $\omega$B97X-V.\cite{wB97X-V} In the latter, the nonlocal dispersion correction of the VV10 type\cite{vv10} was used in its full-SCF implementation; see Ref. \citen{Najibi2018} for a discussion on how VV10 can be applied for this and related functionals. We also note that we demonstrated in our previous study, how the VV10 correction itself contributed to the SIE;\cite{Lonsdale2020} this aspect of $\omega$B97X-V will therefore not be discussed again. Note we only used SVWN5, BLYP, PBE, and B97M-V for Section \ref{Sec:HighZ}.

Higher-orbital calculations for Section \ref{sec:higherangle} were achieved with the maximum overlap method (MOM).\cite{Gilbert2008} 
In this case, a set of ground state orbitals based on the DFA of choice formed the basis for an MOM calculation. This procedure was adopted to calculate the total energies of mononuclear one-electron systems containing the nuclei from hydrogen to carbon when the single electron occupies the 2s or 2p orbitals. An initial calculation with the DFA of choice was done to generate the orbitals, then the occupied orbital was swapped with the higher lying orbital of choice. Following this, an MOM calculation was started with overlap being maximized with the initial guess orbitals.
DFAs tested in this part were: SVWN5, BLYP, PBE, revPBE,\cite{revpbe} RPBE,\cite{rpbe} TPSS, SCAN,\cite{scan} B3LYP, PBE0, M11,\cite{M11} MN15,\cite{MN15} CAM-B3LYP,\cite{CAM-B3LYP} N12-SX,\cite{MN12-SX_N12-SX} $\omega$B97X-V, B2PLYP,\cite{B2PLYP} $\omega$B97M(2).\cite{wB97Mb2b} Note that the last two are double hybrids\cite{B2PLYP,dhdf_wires} and that for one-electron systems, only their hybrid portions are relevant. Again, the choice of DFAs was inspired by our previous findings.\cite{Lonsdale2020} Most of these calculations converged slightly better, which means we could include more DFAs than in our geometry study; exceptions were some gaps for RPBE, TPSS, SCAN, M11, and MN15, for which the MOM procedure did not always converge. Those DFAs were still included where they worked to allow at least for a qualitative study of them.

\section{Results and Discussion}

\subsection{Geometry and the self-interaction error}
\label{sec:geometry_stuff}

In the context of our previous work,\cite{Lonsdale2020} the most logical extension from our original dimer study is to investigate the effect a third nucleus has on the SIE. This is achieved with a heterolytic dissociation curve of the asymmetric, linear H$_3^{2+}$ trimer: the only system with such asymmetry to be considered in our work. Following this, we conduct and analysis of the SIE and its components, specifically of the \textbf{Linear 3} and \textbf{Triangle} geometries. Then, we simplify the discussion by using mean absolute percentage error (MAPE) values to more broadly cover the remaining systems and DFAs studied. In this work, the MAPE is defined as:

\begin{equation}
MAPE = \frac{1}{N} \sum_i^N \left| \left(\frac{SIE^{DFA}_i}{{E_{1el}^{HF}}_i}\right)\right| \times 100 \% \ ,
\label{eqn.MAPE_define}
\end{equation}

which is an averaged sum of SIE values for each distance $i$ for a given system as a percentage of the electronic HF energy, which is identical to $E^{HF}_{1el}$, for that same system and distance. While it would have been desirable to analyze MAPEs for the entire range of $r$ until the HF dissociation limit has been reached, severe convergence issues for $r$ > 10  {\angstrom} for some systems and DFAs prevented us from doing so. For this reason we decided to base all MAPEs on a limit of 10 \angstrom. 

Relevant to later discussion, we substitute eqs \ref{eqn.Cor}, \ref{sie_func_err}, \& \ref{sie_dens_err} into eq. \ref{eqn.MAPE_define} in place of the total SIE to generate MAPEs of various SIE components:

\begin{equation}
MAPE^{corr} = \frac{1}{N} \sum_i^N \left| \left(\frac{{E_C^{SIE}}_i}{{E_{1el}^{HF}}_i}\right)\right| \times 100 \% \ ,
\label{eqn.MAPE_corr}
\end{equation}

\begin{equation}
MAPE^{func} = \frac{1}{N} \sum_i^N \left| \left(\frac{SIE^{func}_i}{{E_{1el}^{HF}}_i}\right)\right| \times 100 \% \ ,
\label{eqn.MAPE_func}
\end{equation}

and 

\begin{equation}
MAPE^{dens} = \frac{1}{N} \sum_i^N \left| \left(\frac{SIE^{dens}_i}{{E_{1el}^{HF}}_i}\right)\right| \times 100 \% \ .
\label{eqn.MAPE_dens}
\end{equation}

As we argued in our previous study,\cite{Lonsdale2020} dissociation curves might cross the x axis for different $r$ values, meaning that different DFAs have a total energy of zero at different internuclear distances. This makes a direct comparison between DFAs difficult, which is why we chose the total HF energy as a common denominator to facilitate a direct comparison between DFAs in dinuclear systems. While this same approach would still work here, as long as DFA trends are analyzed within the same system, preliminary analysis revealed a new difficulty.  Herein, we compare systems that differ in the number of protons while having the same electron number. This means that nuclear repulsion dominates the total HF energies, particularly for shorter values for $r$.  Consequently, systems with more nuclei or denser nuclear arrangements have HF energies with smaller magnitudes in the denominator when calculating the MAPE contributions for shorter $r$, resulting in larger relative MAPEs for these bigger systems. This effect can make up a significant amount of the difference in the MAPE values between systems, with the difference in the SIE coming second to this effect. This might hamper the analysis of SIE trends between systems. For this reason, we decided to only use the electronic HF energy for the present MAPE analysis. As the SIE is not impacted by the nuclear-nuclear repulsion energy either (see Eq. \ref{totalSIE2}), we are therefore able to clearly analyze SIE trends for all DFAs and systems.

MAPE plots will be presented over the entire range of data points, but in addition we focus on a feature of the SIE that is rarely discussed, namely the region of positive SIE for shorter distances. For that reason we carry out an additional analysis where datapoints are separated according to having a positive and a negative SIE, which differs from functional to functional. We also provide an alternative view of our results to complement the MAPE-based analysis, as discussed further below.

\subsubsection{Hydrogen-based dissociation curves}\label{Sec:H-Based-Geom}

\begin{figure}
\begin{center}
\includegraphics[width=1\linewidth]{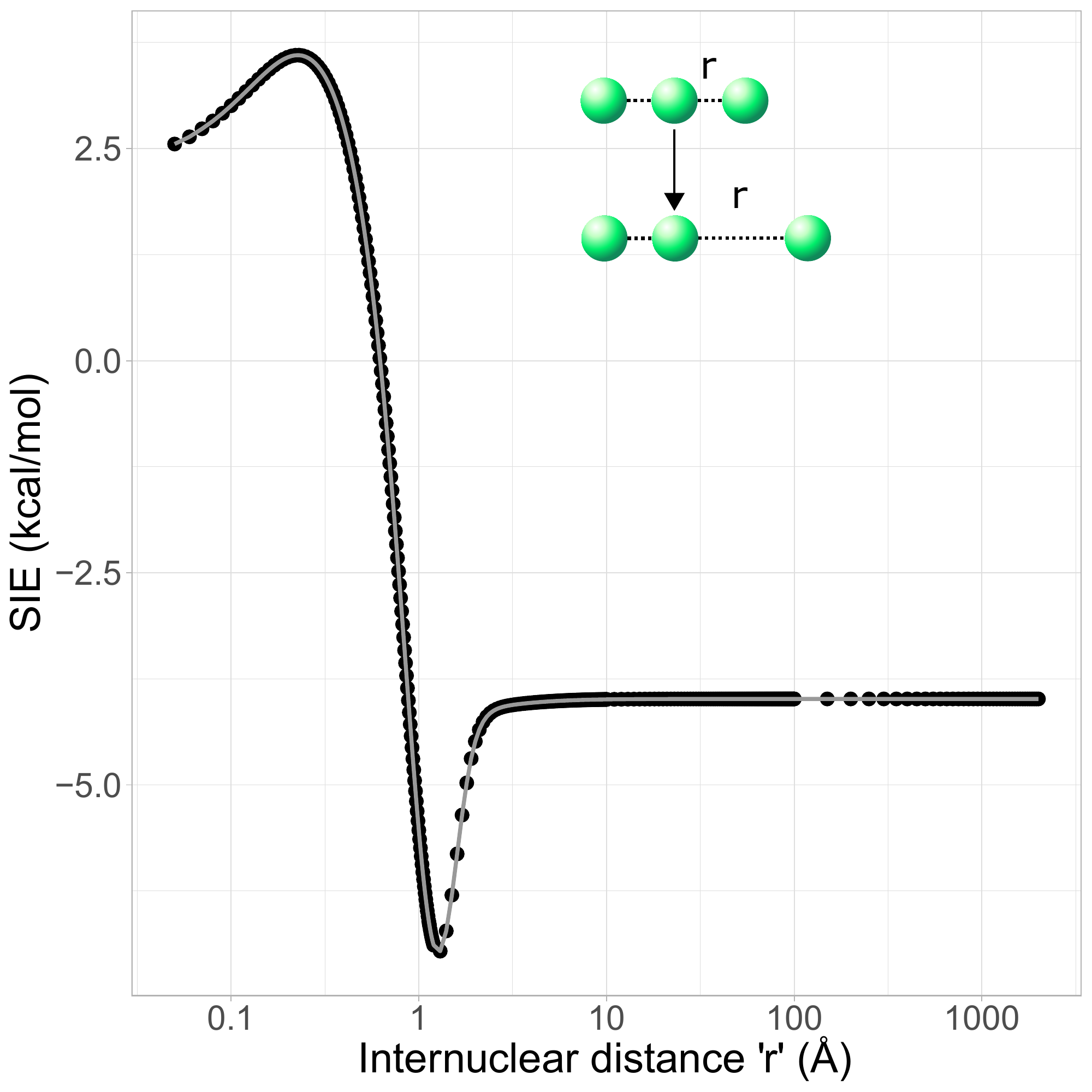}
\end{center}
\caption{H$_3^{2+}$ dissociation. SIE vs the internuclear distance, $r$, which is the separation between H$^+$ and the central nucleus. H$_2^+$ fragment fixed at 1.135 \angstrom. BLYP/uncontracted aug-cc-pVQZ.}
\label{fig:figure2}
\end{figure}

In our previous paper the largest one-electron systems we studied were dinuclear pairs from H$_2^+$ up to C$_2^{11+}$. A natural progression of this is the addition of more nuclei. 
For this section, we exclusively use hydrogen nuclei to explore various one, two, and three-dimensional geometries and their respective SIE behavior.

First, we investigate an analogue to the simple hydrogen abstraction reaction, \ch{H2 + H ->H + H2}, which is known to suffer substantially from the SIE.\cite{H3_barrier_underestimate} To suit the exact one-electron SIE we modified the system to be a dissociation between H$_2^+$ and H$^+$ whereby only one nucleus is moved and the H$_2^+$ fragment is set to have a constant internuclear distance of 1.135 \angstrom, equal to the equilibrium distance in the H$_2^+$ molecule as calculated by our DFA of choice, BLYP.
The resulting dissociation curve is shown in Fig. \ref{fig:figure2}, where we see how the SIE has a strong distance dependence on the third nucleus, with a maximum and minimum occurring before a plateau is reached.
As established by our previous paper,\cite{Lonsdale2020} one-electron systems comprised of nearby nuclei result in positive SIE values: exemplified in Fig. \ref{fig:figure2} when the third nucleus is 0.1 - 0.5 \angstrom\ away from the H$_2^+$ molecule. 
Upon increasing the distance between the fragments a minimum of about $-$7.0 kcal mol$^{-1}$ forms at 1.3 \angstrom\ away from the dimer. Should all three nuclei be allowed to optimize, the delocalization error would be larger in a more symmetric arrangement.
After this minimum, the SIE then converges to $-$4 kcal/mol and does not change thereafter; this constant value represents the SIE of the lone H$_2^+$ ion (for BLYP) and the third nucleus no longer participates energetically, i.e. there is no long-range SIE. This is congruent with our previous findings for the heteronuclear series.\cite{Lonsdale2020}
Therein we dissociated ions such as HHe$^{2+}$ and found no delocalization error, just as we do here with the asymmetric H$_3^{2+}$ system.
To explain this we referred to Perdew and Ruzsinsky that had shown that the typical `delocalization error' manifestation of the SIE is related to electron affinities and ionization potentials of two dissociating fragments.\cite{spurious_FracCharge2006} 
To paraphrase their work, should the ionization potential be similar to electron affinity of the other fragment, then the SIE in the form of the delocalization error will be substantial. While their study was based on chemically different fragments interacting with one another, we can draw analogous conclusions for the present system. When all three hydrogen nuclei are approximately equidistant their ionization potentials and electron affinities are equal. In such cases the DFA overstabilizes the system, the consequence being a more negative SIE. In contrast, the increasingly dissociated dinuclear H$_2^+$ and mononuclear H$^+$ fragments possess different ionization energies  (zero in the case of the latter) and electron affinities. 
These differences explain why the SIE does not continue to decrease into a delocalization-type error that is typical of homonuclear dissociation curves as seen in our previous work.\cite{Lonsdale2020}

Given the above reasoning, asymmetric one-electron model systems at large internuclear separations are expected to produce results analogous to Fig. \ref{fig:figure2}. Therefore, in order to further study the geometrical behavior of the SIE we either need to move to symmetric systems and/or restrict the range of bond lengths studied. To limit complexity to a minimum without compromising any new insights from our study, we decided that the model systems in Fig. \ref{fig:geometries} consist of regular internulear distances $r$, of which \textbf{Linear 3} and the \textbf{Triangle} are the simplest examples.

\begin{figure}
\begin{center}
\includegraphics[width=1\linewidth]{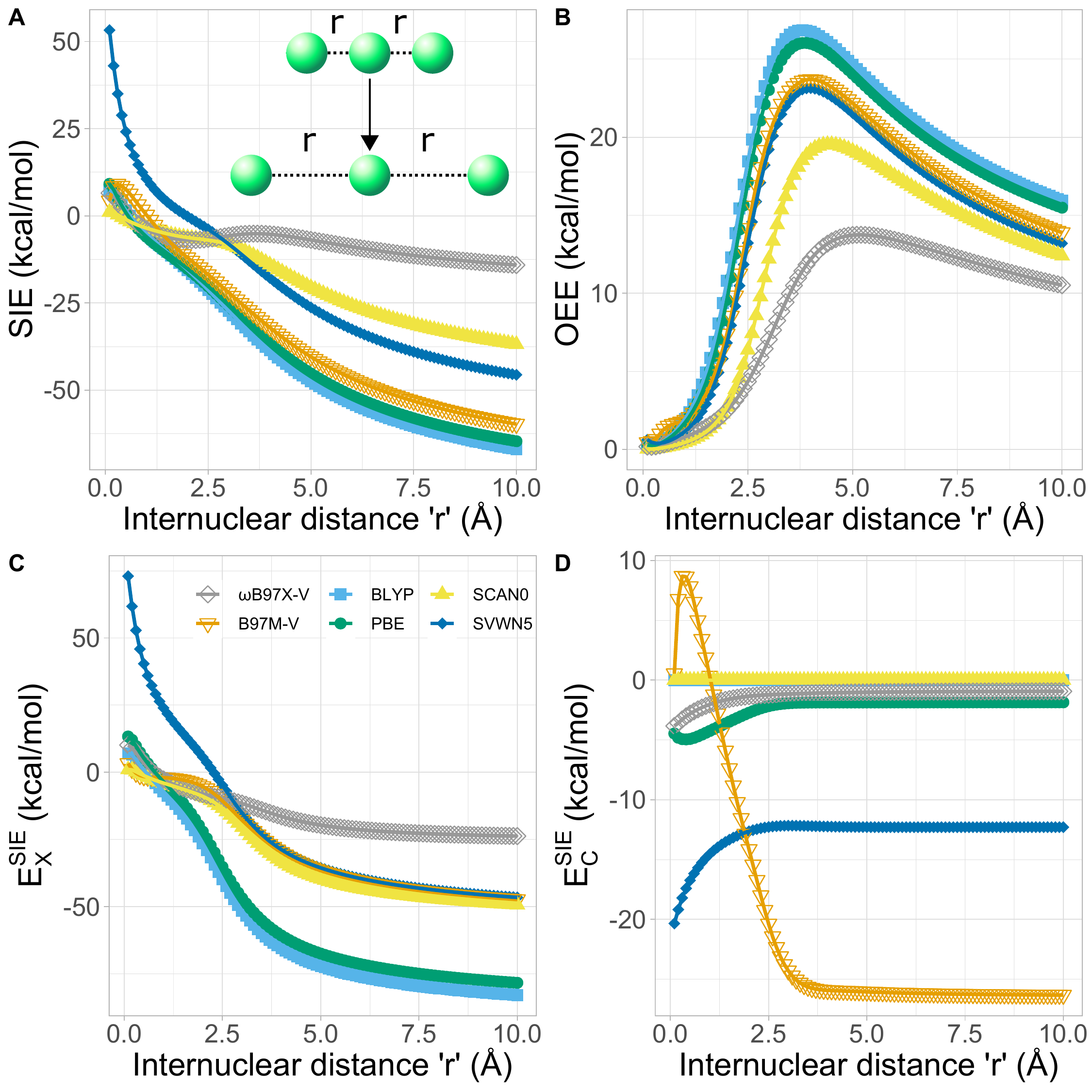}
\end{center}
\caption{Linear H$_3^{2+}$ system showing the self interaction error (A), one-electron error (B), exchange error (C), and correlation error (D) vs distance $r$ between two nearest neighbors.}
\label{fig:figure3}
\end{figure}

\textbf{Linear 3} is the D$_{\infty\text{h}}$ arrangement of the H$_3^{2+}$ ion whereby each nucleus is in a line and each internuclear distance $r$ between two nearest neighbors is identical.
Gradually increasing $r$ results in the dissociation curves shown in Fig. \ref{fig:figure3}.
This four-panel figure shows the SIE, OEE, E$^{SIE}_X$, and the E$^{SIE}_C$, respectively.
Separation into components allows us to clearly see the contributions of each to the shape of the total SIE. The figure is limited to six DFAs, but the results for the remaining five are shown in the Electronic Supporting Information (ESI) Section 1.

Both the magnitude and shape of the total SIE in Fig. \ref{fig:figure3}A are dominated by the exchange error (Fig. \ref{fig:figure3}C). 
All DFAs reproduce this trend, with some DFAs, --- SVWN5 and B97M-V in this case --- also having significant correlation contributions (Fig. \ref{fig:figure3}D). 
At long internuclear distances the SIE continues to grow in magnitude as the lone electron is shared over two or more nuclei, in accordance with the concept of the delocalization error. 
Excluding SVWN5 for now, the GGAs BLYP and PBE have the worst SIEs closely trailed by the mGGA B97M-V, followed by the hybrid SCAN0, and finally the range-separated $\omega$B97X-V.
As expected, range separation reduces the SIE, particularly in the long range, with a sizeable portion remaining at closer internuclear distances --- demonstrating that it is not a complete solution for SIE, which further confirms a similar statement we made in our previous study.\cite{Lonsdale2020}

An important, yet often overlooked, part of the exchange error is the positive component which many DFAs possess in the short range. 
In the cancellation between the exchange and Coulomb energies the former must either be too large or too small compared to the latter. When it is too small, the failure to cancel the Coulomb energy results in the SIE becoming positive --- prominent in SVWN5 in Fig. \ref{fig:figure3}A.
For the H$_2^+$ case we showed that for several DFAs this positive SIE was relevant at the equilibrium bond length for that ion\cite{Lonsdale2020} --- this results in a subtle, but important, distinction. 
If we see a GGA predict a bond length as too long, we might automatically assume that the reason is an over-stabilization from the SIE, rather than the destabilization of the true equilibrium bond length.
As the true SIE is masked in many-electron calculations a negative SIE does not automatically explain all elongated bond lengths, and so one should not assume the sign of the SIE is ubiquitously negative.

We also offer a breakdown of error compensation between the one-electron error, OEE, and the exchange. 
OEE stems from an incorrectly calculated density which causes the one-electron operator to give an incorrect energy. 
This density-based error possesses a substantial positive value, especially for the \textbf{Linear 3} system in Fig. \ref{fig:figure3}B.
In the short range region, $r$ $<1.5$ \angstrom , the exchange error and OEE are both smaller. 
Both errors' magnitudes increase with distance: E$_X^{SIE}$ increases due to the variational method finding a lower total energy through improper cancellation between the self-Coulomb and exchange energies. 
The OEE may be related to this increase in exchange-error --- a larger $E^{SIE}_X$ compounds during the SCF creating an incorrectly calculated density, from which we might expect a larger OEE.
However, the nature of the OEE is not as simple as suggesting that a larger exchange error results in a larger OEE.

 To illustrate the impact of the OEE we give a brief summary of its role in error compensation for the hydrogen atom in PBE. PBE had its $\kappa$ parameter fitted to satisfy the Lieb-Oxford bound,\cite{Lieb1981} and thus yields a very accurate energy for the hydrogen atom. However, the individual exchange and correlation contributions to the total SIE are non-zero.\cite{Lonsdale2020} If we only take error cancellation between the exchange and correlation components we get a total deviation of $-$0.4 kcal/mol. When we also include the OEE in the compensation the deviation is further reduced to a minuscule 0.01 kcal/mol.\cite{Lonsdale2020}
For {\bf Linear 3} we see a similar compensation effect due to the OEE for every tested DFA.

\begin{figure}
\begin{center}
\includegraphics[width=1\linewidth]{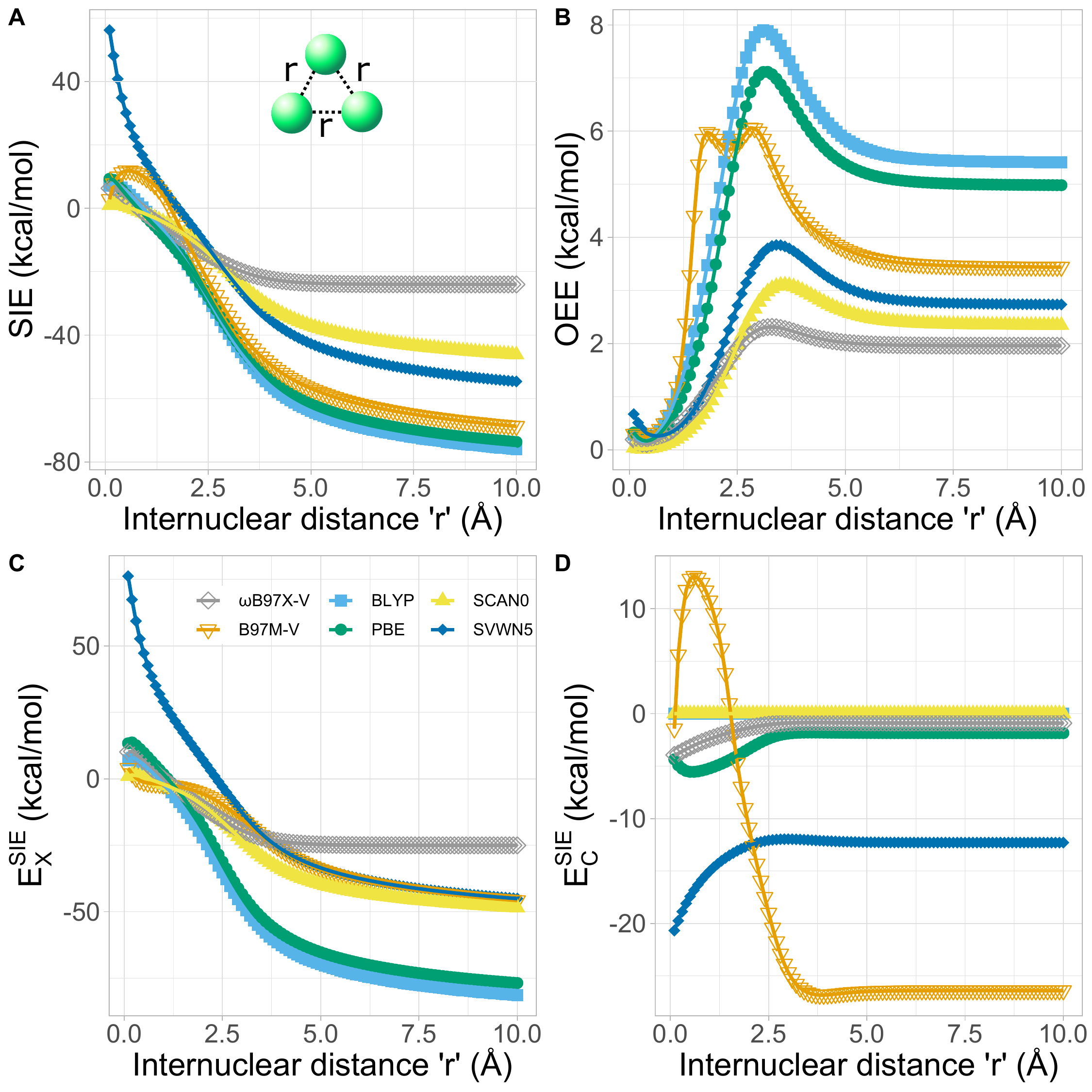}
\end{center}
\caption{Triangular H$_3^{2+}$ system showing the self interaction error (A), one-electron error (B), exchange error (C), and correlation error (D) vs distance $r$ between two nearest neighbors.}
\label{fig:figure4}
\end{figure}

Contrasting the linear system is the \textbf{Triangle}, representative of the two-dimensional geometries, and thus plotted equivalently in Fig. \ref{fig:figure4}: results from all other geometries can be found in the ESI Section 1. 
Akin to \textbf{Linear 3}, Fig. \ref{fig:figure4}A shows that the same typical delocalization error curve is also seen for the \textbf{Triangle}. However, while the SIE continues to slightly grow in magnitude for the linear system, it reaches a plateau  around the 10 \angstrom\ mark for the triangular one.
Moreover, the total SIE at long range is also slightly larger in magnitude for the \textbf{Triangle} and possesses DFA-specific differences in the close range: e.g. B97M-V crosses from having a positive to negative SIE at  approximately 1.0 \angstrom\ for \textbf{Linear 3}, but crosses at about 1.6 \angstrom\ for the \textbf{Triangle}. 
Critically, simply rearranging the geometry of the three nuclei has changed the behavior of the one-electron SIE.

To investigate the cause of the difference between the geometries we will begin with the similarities. Panels D of Figs. \ref{fig:figure3} and \ref{fig:figure4} show the $E_C^{SIE}$ contributions between the two geometries as being nearly indistinguishable in shape and magnitude. B97M-V is the only DFA with much notable difference. Indeed correlation-errors are far less sensitive to geometry in our one-electron test systems in general (see ESI Section 1, E$^{SIE}_C$ values do not change substantially upon change of geometry e.g. Figures S11 and S21, the \textbf{Square} and \textbf{Octahedron}, respectively).

Panels C of both figures show that the exchange-error is again, extraordinarily similar. Especially when approaching 10 \angstrom\ there is an astonishing resemblance between \textbf{Linear 3} and the \textbf{Triangle}, with only a few kcal/mol separating the two. Only very minor differences in the shapes and magnitude of the various DFAs exist.  One such difference in SVWN5 and B97M-V cause the \textbf{Triangle} to have a slightly larger $E_X^{SIE}$ in the short range but less in the long range; such observations are DFA specific.

Finally in panels B of Figs. \ref{fig:figure3} and \ref{fig:figure4} the OEE is the most obviously different in shape and magnitude. 
In particular, the \textbf{Triangle} has an OEE roughly three times smaller than that of the \textbf{Linear 3} system. 
Reduction to the magnitude of the OEE means less error compensation between the OEE and the exchange error is present in the \textbf{Triangle}, which in turn means a greater observed SIE: this represents the largest difference between these two systems. 
Additionally, the OEE in the \textbf{Triangle} tends to become constant quite quickly after it reaches a maximum, this gives the \textbf{Triangle} an overall flatter SIE curve than the \textbf{Linear 3} system. An additional feature is the odd double hump shape of B97M-V in the \textbf{Triangle} (Fig. \ref{fig:figure4}B) hinting at problems that are presumably connected to the highly empirical nature of this functional, as the other assessed DFAs behave more consistently. 

\begin{figure}
\begin{center}
 \includegraphics[width=1\linewidth]{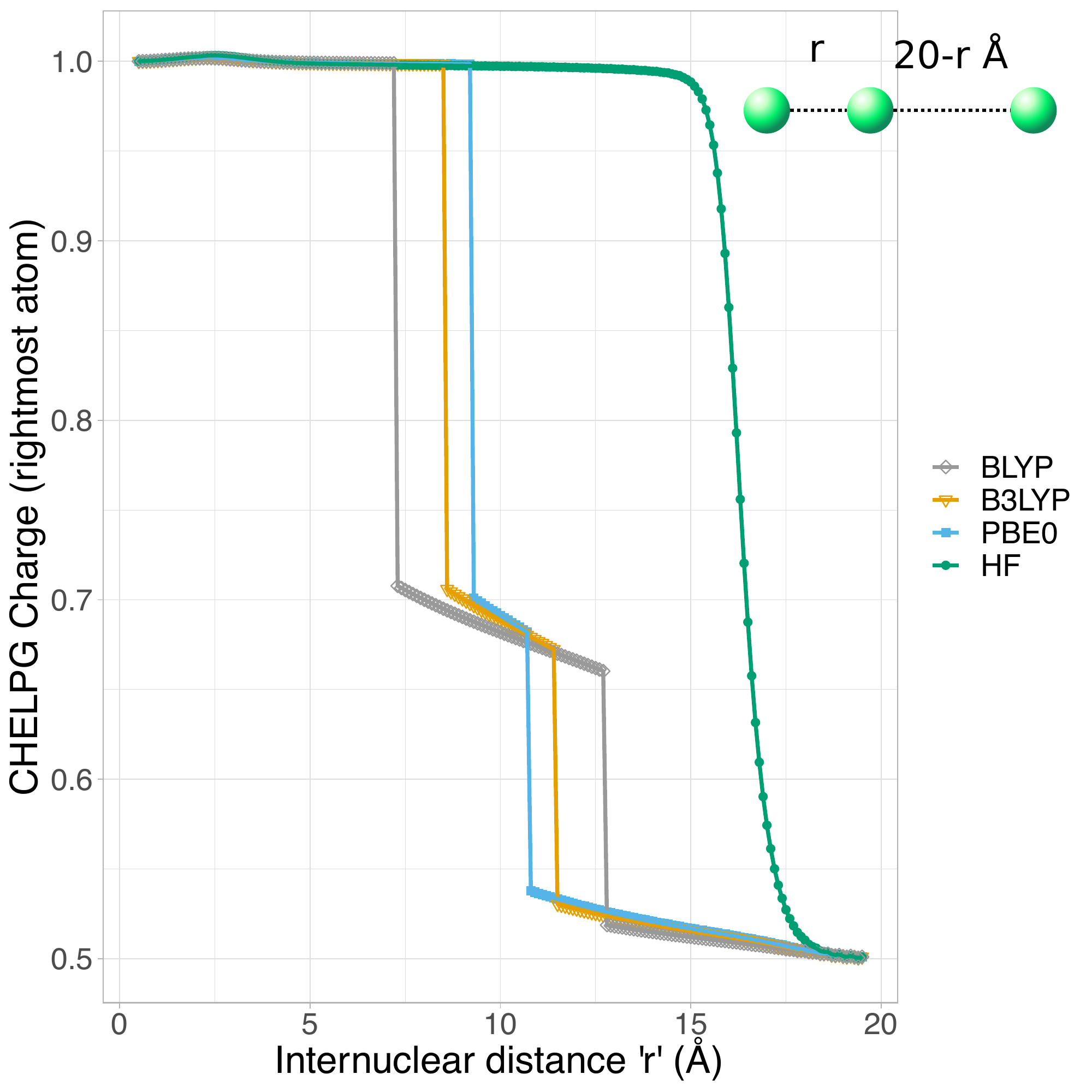}
 \end{center}
\caption{Changes in CHELPG charge on the rightmost hydrogen nucleus in linear H$_3^{2+}$ when observing the migration of the central nucleus from left to right.}
\label{fig:figure5}
\end{figure}

Upon a reviewer’s suggestion, we also provide a brief fractional charge view of the SIE. Such fractional charge analyses are popular in SIE literature. That being said, we conducted such an analysis on mononuclear systems in our previous study in detail where we showed how the conventional fractional charge and our energy based views of one-electron systems produced similar trends and conclusions.\cite{Lonsdale2020} To avoid repetition with little new insight, we restrict our present discussion to {\bf Linear 3} only and focus on the analysis of atomic partial charges rather than setting the charge of the entire system to a fractional one. We created a {\bf Linear 3} geometry with a fixed distance of 20 {\AA} between the two outer nuclei and then gradually moved the central proton from the left to the right with the variable $r$ indicating the distance between the left and central nuclei (see Fig. \ref{fig:figure5}). During this scan we calculated CHELPG\cite{CHELPG} charges, but we would like to note that Mulliken charges gave us the same picture. We show the partial atomic charge on the rightmost atom in Fig. \ref{fig:figure5}.  As expected, HF allocated a +1.0 charge for most values of $r$ until the central nucleus moves into the vicinity of the third (about 5 {\AA} distance between the two) and the charge smoothly decreases to +0.5. This represents the expected finding that the central and right nuclei have started to form the $H_2^+$ molecule. 

For three tested DFAs with different amounts of exact exchange (BLYP, B3LYP and PBE0) we see spurious transfer of fractional charge in two, early-appearing discontinuous steps as opposed to the relatively late, smooth change observed for HF (Fig. \ref{fig:figure5}). For instance, the B3LYP charge on the right nucleus drops from the correct value of +1.0 to about +0.7 at around $r$ = 8.6 {\AA}. This means about 0.3 of an electron got transferred to the right nucleus even though the distance between the right and central nuclei is still too large to expect any electron sharing. A further, smaller jump to about  +0.5 is observed at around $r$ = 11.5 {\AA}. Similar spurious transfers of about 0.3 and 0.2 electrons are also seen for the other tested DFAs at slightly different distances. This example nicely demonstrates the delocalization error problem from a fractional charge perspective.

In this section we have discussed different arrangements of three protons in our one-electron example models and demonstrated how the geometry does indeed have differences on the total SIE and its individual components. Next, we extend the discussion to systems with more nuclei. 

\begin{figure*}
\begin{center}
\includegraphics[width=1\linewidth]{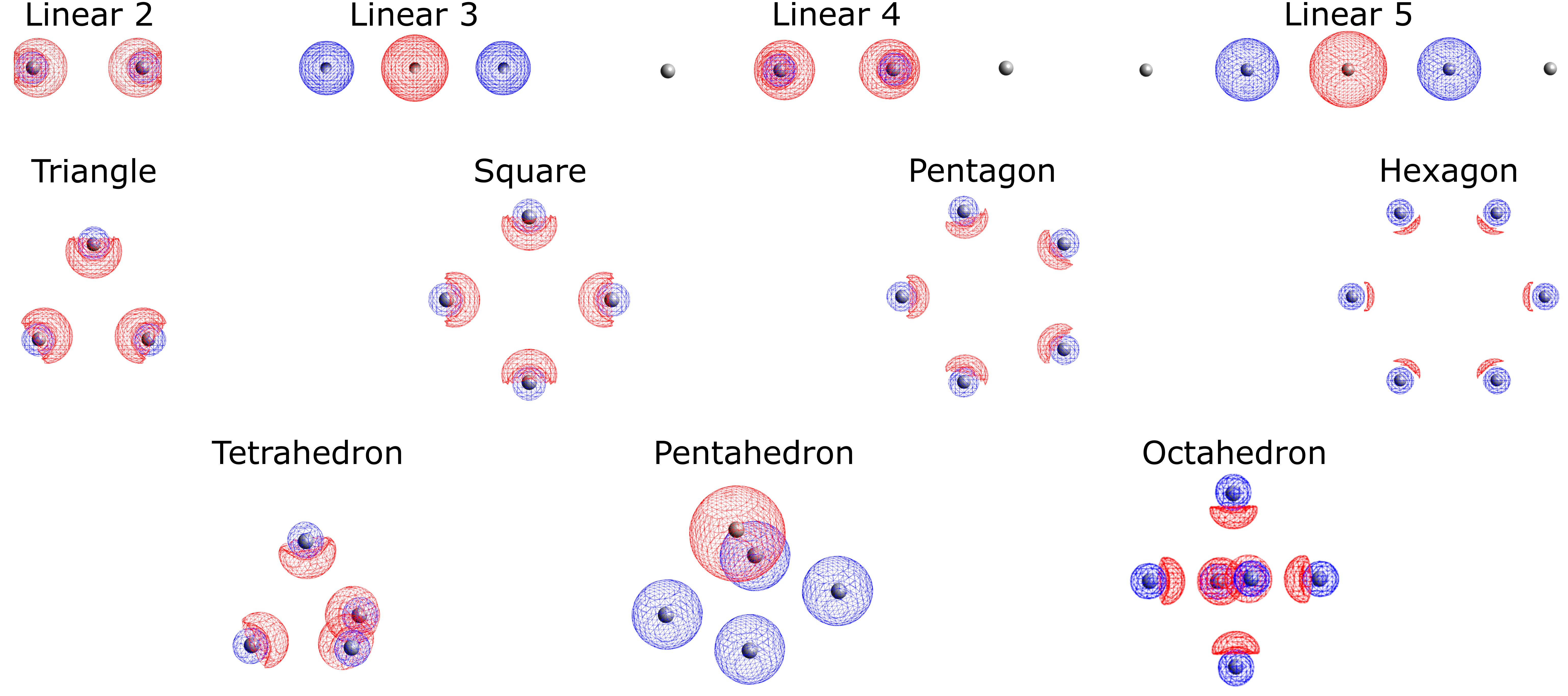}
\end{center}
\caption{Density difference plots between PBE and HF at an internuclear separation $r$ = 5.0 \angstrom\ (as defined in Fig. \ref{fig:figure3}). An isosurface value of 0.0002 e/bohr$^3$ was chosen. Blue indicates a positive difference, red a negative.}
\label{fig:figure6}
\end{figure*}

\subsubsection{Density Difference}\label{Sec:DensDiff}

Before we continue with a quantitative analysis of the observed SIE, we explore a qualitative route first via visualization of the SIE. Such an analysis is an attractive prospect as an intuitive educational tool and a probe to further the understanding of the error and its consequences. With the densities generated from DFA and HF calculations we created the difference densities for every tested geometry for PBE in Fig. \ref{fig:figure6}.
As the SIE is only a fraction of the total energy, correspondingly these densities were only a fraction of the volume of the total density and the isosurface value has been increased to compensate. Regions that have an increased DFA density compared to HF due to SIE are shown in blue and decreases are shown in red. 

For {\bf Linear 3} there is an increase in electron density on the outer nuclei and a decrease to the inner nucleus due to SIE.
In this particular case, the electron is disproportionately localized on the central nucleus for the exact density, however in the case of PBE the electron is located on the outer nuclei slightly more than with respect to the exact solution. 
{\bf Linear 4} shows a different trend, while {\bf Linear 5} resembles the trinuclear system more (Fig. \ref{fig:figure6}). We will see later if this difference can also be reflected in our MAPE analysis.

The symmetry of the \textbf{Triangle} is also reflected in the difference density plot. There is more electron density immediately around each nucleus and less in the regions that lie between the nuclei.
Extending the isosurface would show that there is less density everywhere around each nucleus, except for the blue region immediately nearby, similar to \textbf{Linear 4}.

The triangular arrangement induces equal sharing of the electron by all three centers for both HF and DFT, akin to the symmetry considerations of H$_2^+$.
This is true of every non-linear system assessed in this study except for the pentahedron, where a decrease in density is observed for one of the five nuclei and an evenly distributed increase across the remaining four (Fig. \ref{fig:figure6}).

Our analysis reveals two distinct modes of electron sharing at the HF level: one in which the electron is mostly localized on a single nucleus and one in which the electron is shared across multiple nuclei equally. \textbf{Linear 3} is an example of the former and the \textbf{Triangle} is an example of the latter. Visualizing these highly symmetric one-electron systems demonstrates how the SIE can impact calculations on similar structures. 
\textbf{Linear 3} is representative of the transition state of a hydrogen-transfer or, more broadly, even an S$_N2$ reaction; in such instances, the electron sharing is more widely spread compared to HF and would overstabilize the transition state at the approximate DFT level.
The \textbf{Hexagon} shows an increased stability of an electron throughout the ring, the \textbf{Octahedron} reproduces quite well ligands of a metal complex sharing one-electron over the 6 coordinating nuclei.
While it is another question as to exactly how much one-electron character is inherited in many-electron systems, we believe that there are reasonable grounds to assert that the one-electron SIE is (situationally) substantial based on the success of the Perdew-Zunger correction in select cases.\cite{Ruzsinszky2008,Patchkovskii2002,Klpfel2012}

The difference density shown here is a convenient tool to qualitatively analyze the effect of the one-electron SIE. As a pedagogical tool we hope similar visualizations can communicate the effects of the various errors plaguing DFT, perhaps in conjunction with benchmarking data.

\subsubsection{Hydrogen-based MAPE}\label{Sec:H-Based-Geom-MAPE}

\begin{figure*}
\begin{center}
\includegraphics[width=1\linewidth]{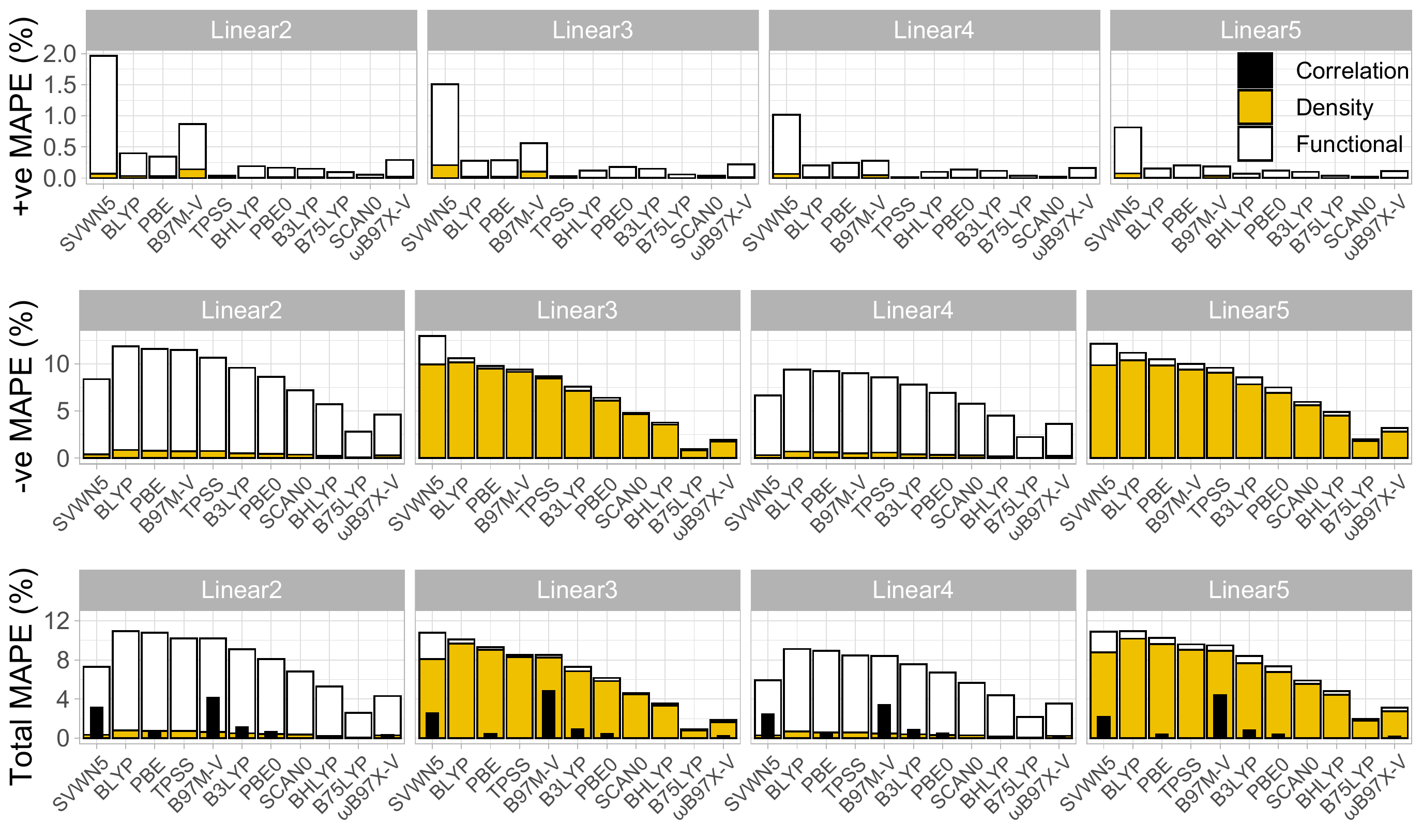}
\end{center}
\caption{Mean absolute percentage error in linear systems broken into positive SIE values (top), negative SIE values (middle) and the total SIE (bottom). 
Functional error (white) and density error (yellow) is presented as a stacked bar plot with their combined heights corresponding to the total MAPEs for the positive, negative, and total SIEs respectively. The bottom row also shows the correlation MAPE (black) separately.}
\label{fig:figure7}
\end{figure*}

MAPE values for the one-dimensional geometries {\bf Linear 2} to {\bf Linear 5} are presented in Fig. \ref{fig:figure7}. We have presented the density  (eq. \ref{eqn.MAPE_dens}) and functional (eq.  \ref{eqn.MAPE_func}) MAPEs as a stacked bar plot, meaning that their combined height is equivalent to the total MAPE (eq. \ref{eqn.MAPE_define}). The top row of  Fig. \ref{fig:figure7} does this for the positive SIE region and the middle row for the negative SIE region (see above discussion of MAPEs). The bottom row shows the MAPEs for all data points in addition to the correlation MAPE (eq. \ref{eqn.MAPE_corr}), which is shown as a separate, independent bar. The plot is ordered according to the rungs of Jacob's ladder---with the additional distinction that range-separated hybrids follow global ones---then in terms of overall performance, namely SVWN5, BLYP, PBE, TPSS, B97M-V, B3LYP, PBE0, SCAN0, BHLYP, B75LYP, and $\omega$B97X-V. In the following discussion, we will use ``MAPE'' and ``SIE'' as synonyms as they clearly correlate with one another.

\begin{figure*}
\begin{center}
\includegraphics[width=1\linewidth]{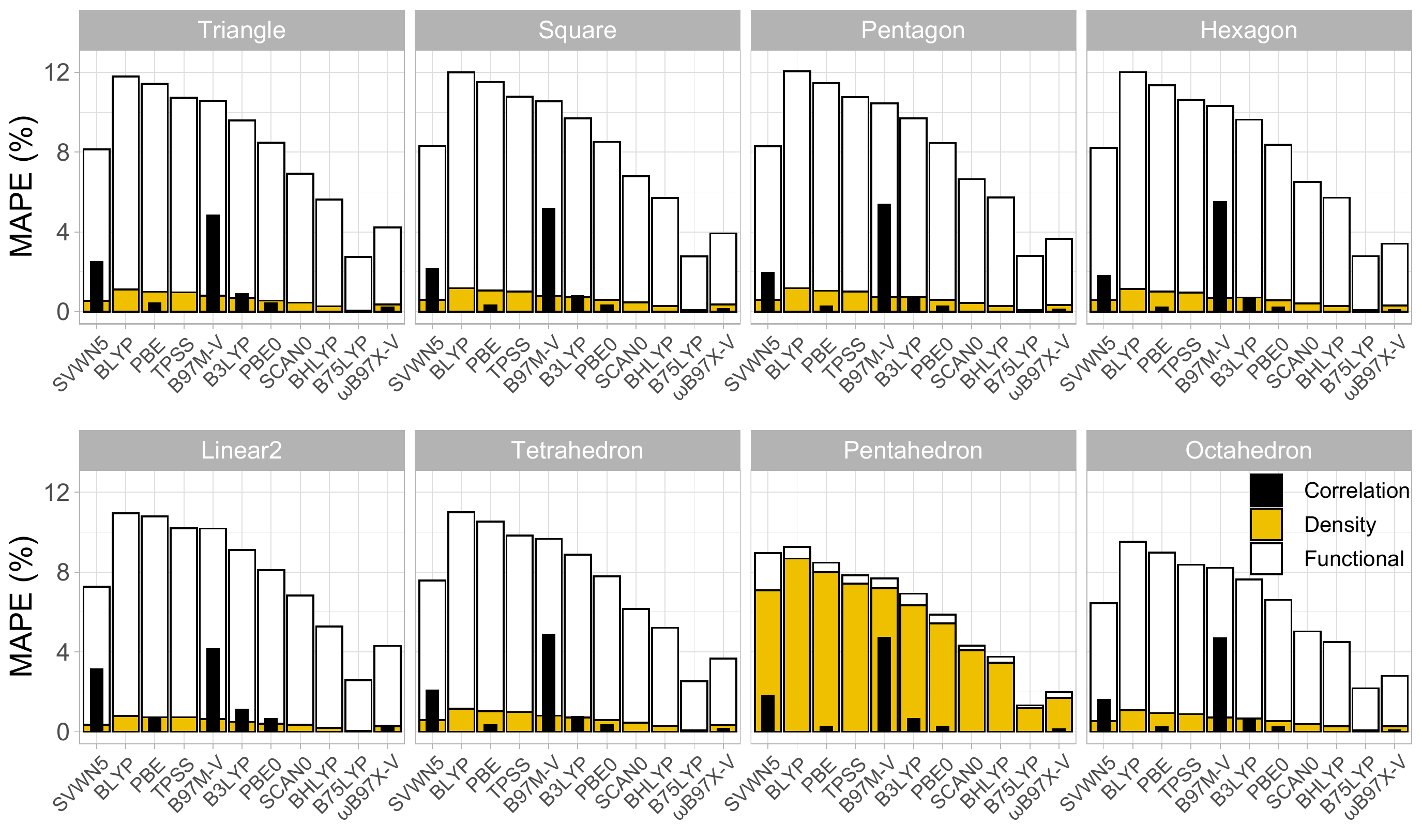}
\end{center}
\caption{Mean absolute percentage error in 2D (top) and 3D (bottom) geometries for all datapoints (positive and negative SIE). Total MAPE is equal to the combined height of the functional (white) and density error (yellow) with correlation (black) shown separately.}
\label{fig:figure8}
\end{figure*}

Comparing each of these systems in order, we first analyze \textbf{Linear 2} and \textbf{3}. The functional rankings in total MAPEs across all $r$ up to 10 {\AA} are very similar for both systems with the exception of SVWN, which ranks as the worst in {\bf Linear 3} but is similar to some hybrids for {\bf Linear 2} (Fig. \ref{fig:figure7}). TPSS and B97M-V have very similar total MAPEs. B75LYP is, unsurprisingly, the best DFA for both cases, followed by $\omega$B97X-V. Some MAPE values change dramatically for certain DFAs, for instance B75LYP sees a reduction of more than half in \textbf{Linear 3} compared to \textbf{Linear 2}, while $\omega$B97X-V sees only marginal reductions. We can see that the addition of the third nucleus has a DFA-specific impact on the total MAPEs. Closer inspection of the individual components of the MAPE, however, reveal distinct differences between both systems and we see a large increase in the density error of {\bf Linear 3}. We can relate this directly to the OEE, which shows a near 10-fold increase for \textbf{Linear 3}. \textbf{Linear 3} also has a larger exchange error, but the increase of this is offset by the increase in the OEE, and thus results in mostly marginal changes to the total SIE compared to \textbf{Linear 2} (Fig. S25).

An interesting picture emerges when separating datapoints that represent positive SIEs from those with negative. While the exact crossover point from positive to negative differs from DFA to DFA, we see for all DFAs and linear systems that the functional error is the dominating contributor to the positive SIE (top row of Fig. \ref{fig:figure7}). This picture changes when looking at regions that only incorporate negative SIE where we see that {\bf Linear 3} and {\bf Linear 5} become dominated by the density error (middle row of Fig. \ref{fig:figure7}). The negative region therefore impacts the total MAPEs more (bottom row of Fig. \ref{fig:figure7}), which is expected as the total MAPE is comprised mostly of datapoints in the negative-SIE region. We also note that in \textbf{Linear 4}, the single electron is delocalized over two nuclei and not four for larger $r$, which means this system most closely resembles H$_2^+$ ({\bf Linear 2}) in terms of delocalization error, which reflects our discussion of difference densities in Section \ref{Sec:DensDiff} and further demonstrates the usefulness of the above introduced visualization. 

When a system's SIE is comprised mostly of density error, as \textbf{Linear 3} is, a large amount of the SIE from that system could be resolved by density-corrected DFT as proposed by Burke and Sim.\cite{Sim2014}
While we have not shown it in this breakdown, the density error has a large component that stems from the exchange part of the functional itself --- in such cases the OEE represents only a small part of the SIE$^{dens}$. 
In fact, for B75LYP and $\omega$B97X-V the \textbf{Linear 3} system has a smaller SIE than the dinuclear counterpart, this is likely owing to the high proportion of exact exchange in these DFAs having a greater effect on reducing the density-based error. This explains the aforementioned drop in total MAPE. The top row in  Fig. \ref{fig:figure7}, however, indicates that DC-DFT might not be helpful to address the positive SIE region.

Just as with the linear test cases, Fig. \ref{fig:figure8} shows the MAPE values of each of our chosen two- and three-dimensional shapes over all datapoints; because the scales are the same, \textbf{Linear 2} is included in the 3D geometries row as a point of reference. For the 2D cases, the total MAPEs remain surprisingly similar to the linear systems with some differences present. Again, the correlation error usually plays a relatively small role, with SVWN5 and B97M-V being exceptions, and it is instead the exchange error that is by far the largest contributor to the observed MAPEs (Fig. S24 for 2D). Comparison of the \textbf{Triangle} to \textbf{Linear 2} (Figs S24 and S25) reveal that the exchange error grows in magnitude upon the addition of the third nucleus. We also observe that all tested two-dimensional structures equally share their electron over all nuclei at both HF and DFA levels of theory resulting in SIE that is dominated by the functional error. Fig. S26 in the ESI shows that the functional error dominates both the short-range region with positive and the long-range region with negative SIEs. In passing we note that the OEE does not impact the MAPEs for the 2D systems, but for specific internuclear distances it can be sizeable and partially cancel out the exchange error (Figs S9 - S16).

The three-dimensional geometries mimic common nuclear arrangements in molecules and complexes (Fig. \ref{fig:figure8}). With the exception of the {\bf Pentahedron} all tested 3D structures are dominated by the functional error. In the case of the  {\bf Pentahedron} the total MAPE values are nearly identical to the density-error MAPE values. Similar to \textbf{Linear 3} and \textbf{5}, only the region with negative SIE is dominated by the density error for this system (Fig. S27) Overall, there is a small reduction in the MAPEs when going from the 2D to the 3D systems. Similarly to the 2D systems, the exchange error dominates in all 3D systems, while OEE-based MAPE is nearly negligible except for the {\bf Pentahedron} (Fig. S25).

For all systems, we observe the trivial result that high amounts of Fock exchange can drastically improve density-error dominated systems, see B75LYP and $\omega$B97X-V in the case of \textbf{Linear 3}, \textbf{5} (Fig. \ref{fig:figure7}) and the \textbf{Pentagon} (Fig. \ref{fig:figure8}). Similarly, relative to its performance in functional-error dominated systems SVWN5 performs worse for density-error dominated systems, chiefly \textbf{Linear 3} and \textbf{5} (Fig. \ref{fig:figure7}), and the \textbf{Pentahedron} (Fig. \ref{fig:figure8}).

\begin{figure}
\begin{center}
\includegraphics[width=1\linewidth]{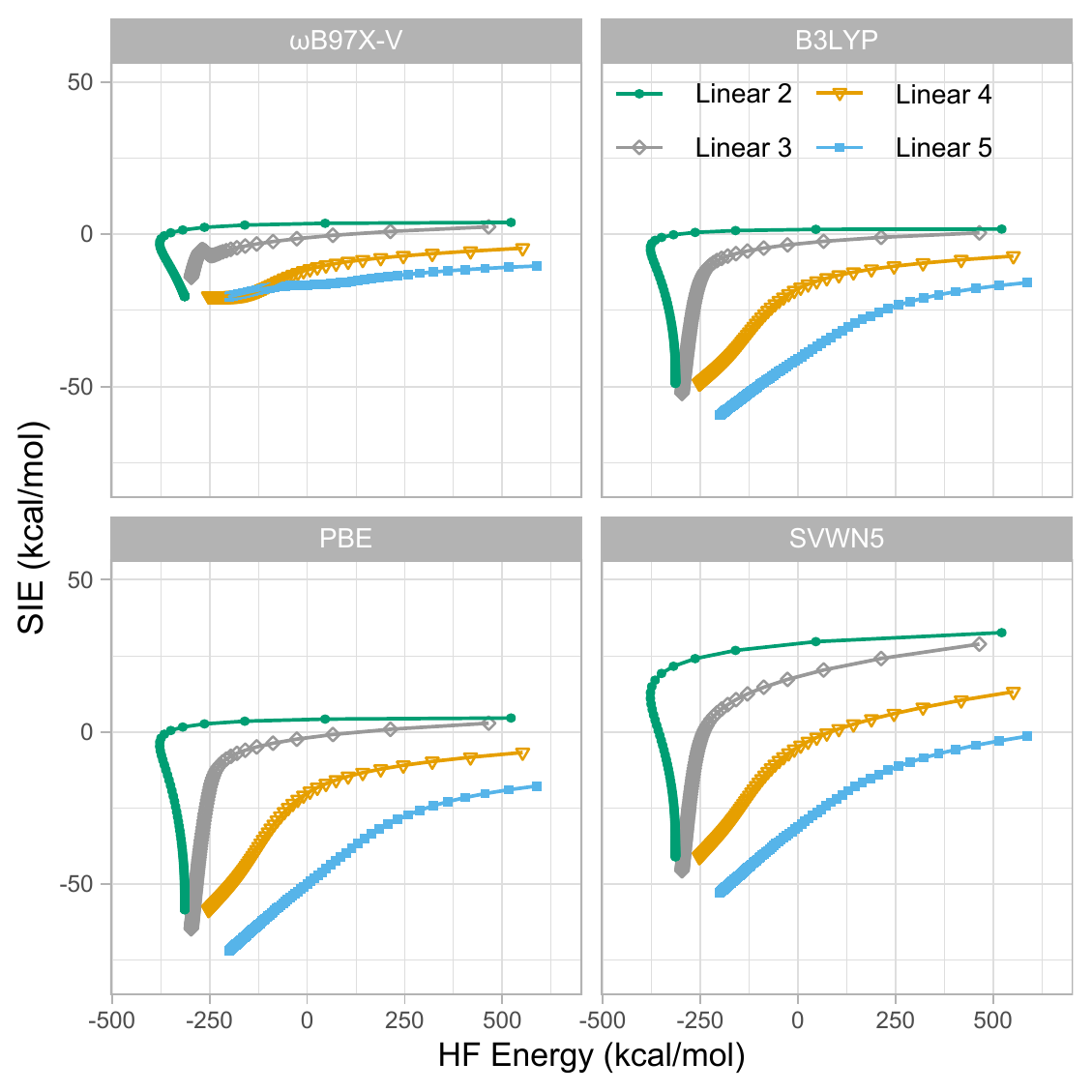}
\end{center}
\caption{SIE vs total HF energy for different internuclear separations, $r$, for the one-dimensional systems.}
\label{fig:figure9}
\end{figure}

While the MAPE analysis allows us to analyze the SIE depending on its sign and its components, a direct comparison between MAPEs for different systems is not possible with the data we were able to obtain. In the dissociation limit, all model systems will converge to a total energy of $-$0.5 $E_h$ at the HF complete basis set level of theory. However, none of these highly charged cationic systems have converged to this limit by 10 {\angstrom} and due to convergence issues we cannot reliably capture the SIE at larger distances. When comparing these different geometric shapes, the HF energies of the different geometries can be quite different at the same distance. We therefore present a complementary analysis in Fig. \ref{fig:figure9} to demonstrate the same information from a different angle. Fig. \ref{fig:figure9} shows the SIE plotted against the HF total energy of each system. It demonstrates how the SIE is substantially different for the different model systems, more so than the MAPE analysis would make us believe. For instance, {\bf Linear 3} has less SIE than {\bf Linear 4}, which in turn has less than {\bf Linear 5} at similar HF energies. \textbf{Linear 2} and \textbf{3} are closer to the dissociation limit by 10 \angstrom\ and yet the SIE continues to decrease despite little to no change in the HF total energy, however it is difficult to say whether each system's SIE will converge to different values in the long range. Increasing SIEs with system size have previously been reported for ionization potentials of many-electron systems,\cite{SIEIPmanyelectron} and here we establish a similar observation for the one-electron SIE. Note that the reason for the odd appearance of the graph for \textbf{Linear 2} compared to the other systems is because \textbf{Linear 2} forms a stable energy minimum, meaning that we can get two different SIE values for similar HF total energies. The other systems do not show this behavior. Another caveat is that the density differences plotted in Fig. \ref{fig:figure6} showed \textbf{Linear 2} and \textbf{4} being almost identical and yet there is a very discernible difference in Fig. \ref{fig:figure9} --- the same story as \textbf{Linear 3} with \textbf{Linear 5}. This yet again shows how sensitive the SIE is to geometry, even in cases where difference-density analysis might lead us to believe that larger systems are extremely similar to the subunits they are built upon. This finding may have consequences to the conclusions of benchmarking studies when extrapolating results to systems outside the scope of the present study.

To offer a robust discussion of the SIE another key feature to track for the discussed systems is the crossover point,  i.e. the distance value in {\angstrom} when the total DFA energy goes from positive to negative compared to when the total HF value crosses this ``zero line''. Given the general shape of these curves it follows that the larger this difference is, the more impactful the SIE value will be to the total DFA energy. 
If we just consider the total energy curve of our dissociating ions for a moment, their shape goes from a region of high repulsion and to a region of negative energy with the total energy of a single hydrogen atom in the exact limit. This means the dissociation curve must cross the zero line at some key distance. Approximate methods will deviate from this curve and the difference of the deviations along the length of the curve constitute the SIE, however the DFA and HF dissociation curves have shapes and gradients that are very similar.
Therefore, the difference between where a DFA and HF predict the change in the total energy from being positive to negative, crossing the zero line, can be used as an indicator to the SIE.
For example, in the \textbf{Octahedron}, BLYP's crossover point is at 7.9 \angstrom\ whereas for the exact HF result the crossover is at 10.0 \angstrom. 
Due to the SIE alone the prediction of a negative-energy structure emerges \textapproxtilde 2.1 \angstrom\ earlier than it should have.
This secondary characteristic can be used to infer the SIE behavior of each DFA for a given system, just like the MAPE. We have aggregated these crossover distances in Fig. S28. A greater number of nuclei and more symmetric shapes (e.g. a tetrahedron having a higher symmetry than a square) have greater crossover errors. Therefore, it is inferred that despite the MAPE analysis prior, the SIE is larger for systems with more nuclei and those of higher symmetry. In conjunction with our analysis of Fig. \ref{fig:figure9} we believe there is sufficient evidence that it is likely the SIE is larger when a single electron is delocalized over more than two nuclei. 

For all systems and most DFAs, the correlation error makes up only a small portion of the total SIE. The correlation errors produced by the individual DFAs maintain an ordering amongst themselves: for instance, B3LYP's correlation MAPE is always larger than PBE's and that of $\omega$B97X-V is smaller than both. This speaks of a relatively predictable nature of correlation-induced SIE in such one-electron systems. For a given DFA, the correlation MAPEs are relatively constant for all systems given the small fraction they make up to the total MAPE. B97M-V and SVWN5 are the only exceptions and they possess significant correlation. Given their larger correlation MAPE, we observe some system-dependent fluctuations for them. For instance, the smallest correlation MAPE for B97M-V is 3.4 \% for \textbf{Linear 4} and the largest is 5.5 \% for the \textbf{Hexagon}.

\subsubsection{Higher nuclear charge}\label{Sec:HighZ}

\begin{figure}
\begin{center}
\includegraphics[width=1\linewidth]{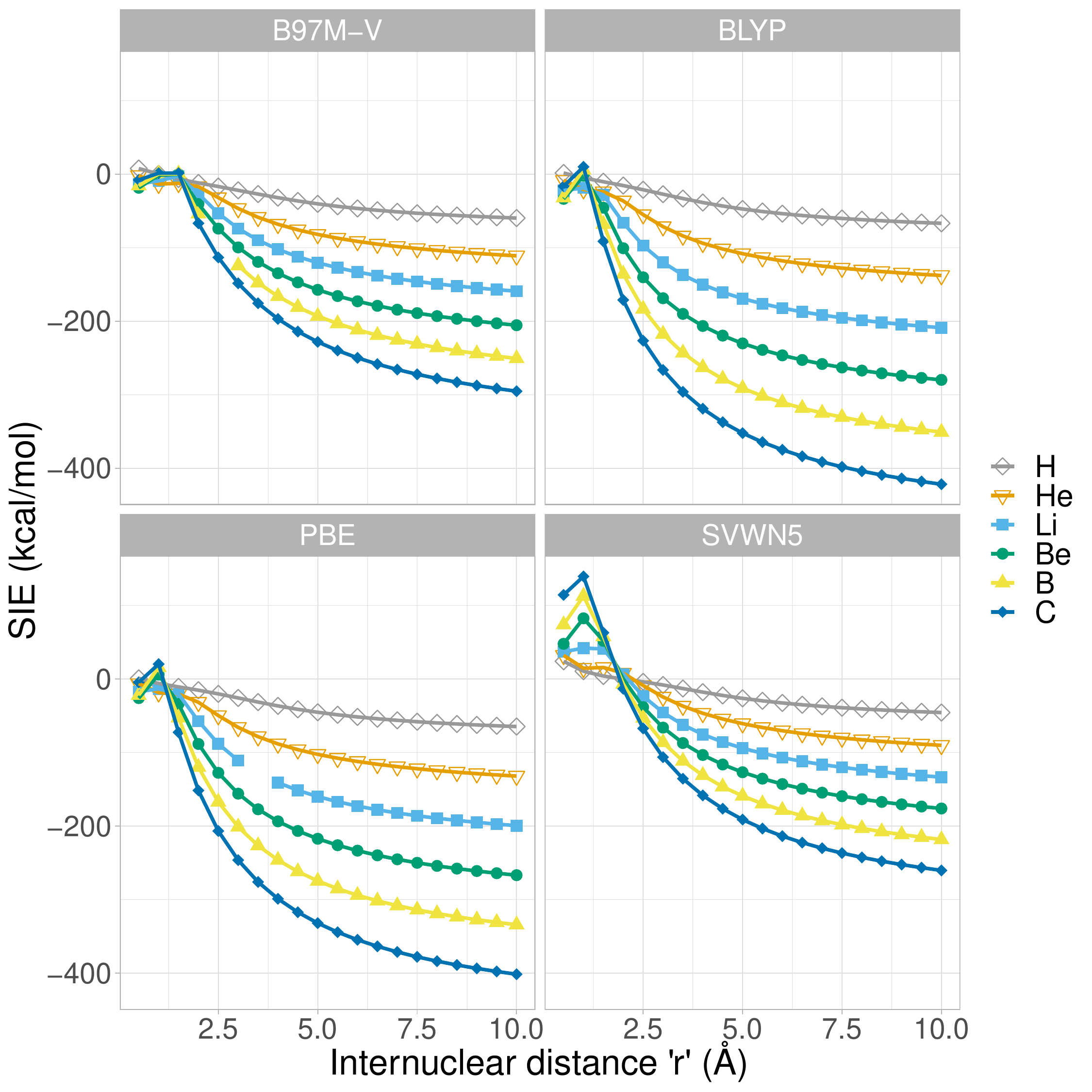}
\end{center}
\caption{SIE vs internuclear distance for increasingly heavier nuclei exemplified for one-electron {\bf Linear 3} and four DFAs.}
\label{fig:figure10}
\end{figure}

As protons were exclusively used in the SIE analyses so far, we now introduce increasingly heavier nuclei from He to C to these same geometries whilst maintaining their one-electron character. 
As a representative case for the higher nuclear charge we have plotted the \textbf{Linear 3} system in Fig. \ref{fig:figure10} to show that increasing the atomic charge $Z$ increases the magnitude of the SIE. In the case of SVWN5 the SIE has a substantial positive value at closer internuclear distances; in order of SIE magnitude, the carbon system's SIE is largest and the hydrogen one's is smallest. Likewise, the carbon system possesses the most negative SIE at elongated distances: generally speaking this is true no matter which DFA is under investigation, some like $\omega$B97X-V may have a different ordering at points due to the range separation procedure.

Even though we only show the \textbf{Linear 3} system this trend is essentially the same in nature no matter which geometry is under investigation (ESI section 5). This finding also reflects our work done on increasing nuclear charge in atomic and dinuclear species in our previous work using a simple argument to explain the linear increase in the SIE --- see Ref. \citen{Lonsdale2020} for more details. This brief example illustrates that this is consistent behavior across many systems and is not limited to the model systems discussed in our last study.

\subsection{Self-interaction error for an electron in higher-lying orbitals}
\label{sec:higherangle}

In all previous calculations we only considered the SIE through ground state calculations of one-electron systems, specifically the 1s orbital or molecular orbitals formed from them. However, if we consider DFT applications to reactions between many-electron systems, SIE effects from valence electrons are expected to be dominating, as SIE in core electrons is expected to cancel between the product and reactant sides. It is therefore reasonable to assume that the SIE based on our hydrogenic systems is only partially representative. To this end we utilized the maximum overlap method (MOM)\cite{Gilbert2008} to force lowest-energy solutions for configurations in which the electron occupies the 2s and 2p orbitals in mononuclear systems. The difference between DFA and HF energy for such solutions is then again defined as the total SIE. At the time we conducted this work, such an analysis of the one-electron SIE and its relationship to orbital occupation had not been conducted before. During the final stages of refining the first version of our manuscript, we became aware of a preprint by Schwalbe, Trepte, and Lehtola --- subsequently published in print while our work was under review.\cite{Lehtola2022} Therein, they calculated the 1s, 2p, and 3d states mononuclear systems from H to Kr$^{35+}$ --- the same systems we used in our previous study.\cite{Lonsdale2020} They used a methodology to calculate the excited states by including only the basis s, p or d basis functions in the calculation, so that the ground state of that calculation would be the 1s, 2p, or 3d respectively. Their study was a thorough ranking/benchmark of many DFAs with a basis set and grid dependence analysis. Our study is more qualitative in nature as we do not seek to rank the DFAs, but just investigate the SIE's effect on higher orbital occupation, and as we used the MOM algorithm we also gained insights into the 2s excited state.

\begin{figure}
\begin{center}
\includegraphics[width=1\linewidth]{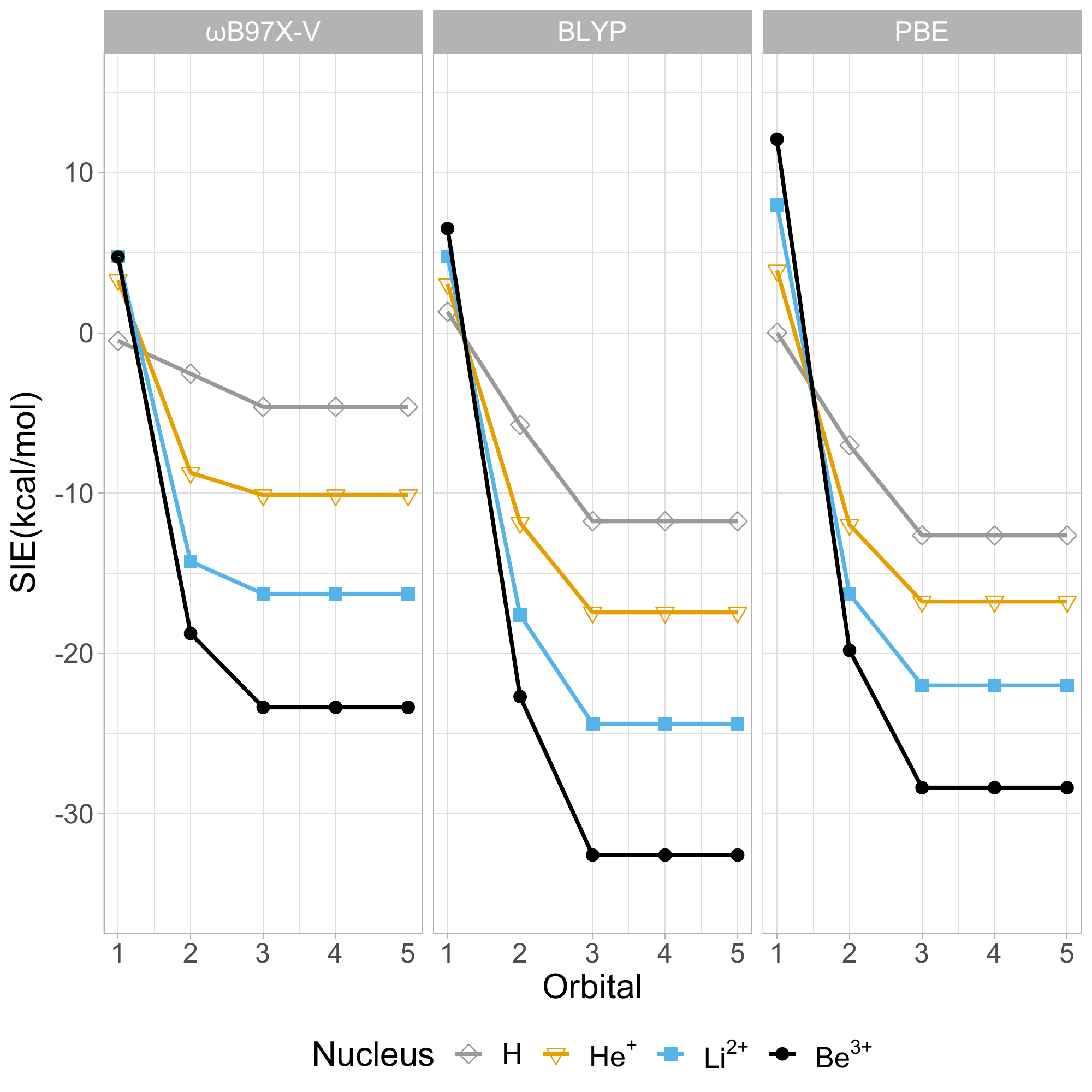}
\end{center}
\caption{SIE vs occupied orbital for increasing nuclear charge for select DFAs. The occupied orbitals are the 1s (1), 2s (2), and 2p (3-5) orbitals.}
\label{fig:orbs3}
\end{figure} 

For the 1s, 2s, and 2p subshells we have plotted the occupied atomic orbital versus the SIE in Fig. \ref{fig:orbs3} for select DFAs; more results are shown in the ESI (Section 6). 
We included the 1s orbital, as it is the ground state for comparison to the higher-lying states. 
For all tested DFAs the SIE becomes increasingly negative with higher orbital occupation, noting that the last three orbitals in Fig. \ref{fig:orbs3} are the degenerate 2p orbitals. 
This general trend can be explained with the size of the 2s and 2p orbitals, or more specifically with their related, well-known radial distribution functions which show that 2s and 2p are more diffuse than 1s. This allows for a greater delocalization of the electron due to SIE. This echoes our finding from the basis set dependence of the SIE where we related more diffuse atomic orbitals with a greater electron delocalization causing a more negative SIE.\cite{Lonsdale2020} 

We already mentioned that PBE has nearly no SIE for the hydrogen atom due to satisfying the Lieb-Oxford bound\cite{Lieb1981} from near perfect error compensation between the exchange, correlation and one-electron errors.
However, Fig. \ref{fig:orbs3} shows that the satisfaction of this condition does not alleviate the SIE of hydrogenic He, Li, or Be, nor does it curtail SIE in higher occupied orbitals. The TPSS and SCAN DFAs are also one-electron SIE free, per construction, but again, this only seems to hold for the 1s orbital of the H atom (see Fig S41).

Viewing Fig. \ref{fig:orbs3} from the perspective of atomic charges shows again that heavier nuclei give more pronounced SIEs. 
From H to Be$^{3+}$ there is again an approximately linear increase in one electron systems --- even when applied to higher orbital occupation. 
Previously in this paper we already found how this trend presented itself in 1, 2, and 3D geometries, as well as in the simple mono- and dinuclear components of our original work.\cite{Lonsdale2020} 
BLYP and PBE show this for the positive SIE in the 1s  ground state as well as for the negative SIEs in the 2s and 2p orbitals. 
An exception exists for $\omega$B97X-V in which the positive SIE for the hydrogenic 1s ground state does not appear to have a neat linear increase for increasingly higher $Z$. 
However, we already showed in the aforementioned work that DFAs with range separation can defy SIE linearity with respect to nuclear charge.\cite{Lonsdale2020}

Among the tested DFAs $\omega$B97X-V performs the best due to it being a range-separated hybrid with 100 \%\ exact exchange in the long range. That being said, we remind the reader that --- in at least the case for these mononuclear tests --- there still is quite a significant value of SIE especially in the heavier nuclei and the 2s and 2p orbitals indicating that range separation is still not a panacea for problems pertaining to self-interaction.

These results demonstrate how fixing the one-electron SIE for one particular state or element does not necessarily guarantee an overall SIE-free functional approximation. The herein found orbital dependence does not only affect the treatment of valence shells in ground-state chemistry, but will most likely also impact any excited-state treatments --- especially in ghost states whereby a higher lying excitation is spuriously lowered beneath the true excited state. Our methodology is in agreement with Schwalbe, Trepte, and Lehtola's\cite{Lehtola2022} comprehensive benchmark of hydrogenic nuclei in the 1s, 2p, and 3d states. Across a larger test set they found that the mean signed errors are in general smaller for the 1s, increase for the 2p, and are largest for the 3d orbital and also had DFA specific performance throughout.

\section{Summary and conlusions}
\label{sec:conclusions}
This work's purpose was to investigate the interplay between the one-electron self-interaction error (SIE) and different geometries as well as higher orbital occupation.
We found a high degree of variability in the behavior of the SIE from seemingly very similar conditions i.e. geometries.
This is testament to the importance of understanding the limitations and behaviors of density functional approximations (DFAs) before applying them blindly to a system.

Our previous study into the one-electron SIE had analyzed 74 popular, well performing, and some of the latest DFAs on one-electron mononuclear and dinuclear model systems with varying nuclear charge.\cite{Lonsdale2020} 
The logical extension of that work was to include a third nucleus into those systems. If done asymmetrically, we observed an impact at closer internuclear distances but none in the long range --- this is to be expected from the work of Perdew and Ruzinsky on heteronuclear diatomics.\cite{spurious_FracCharge2006}  Following that initial extension, we then turned to choosing symmetrically dissociating nuclei arranged in one, two, and three dimensional geometries with up to six nuclei. The aim was to probe the one-electron SIE for any geometry dependence. Indeed, we found a geometry dependence of the SIE. For instance, in systems where the electron delocalized over more than two nuclei, we observed a change in SIE. Differences were also observed between systems of same constituents but different nuclear arrangements, for instance  \textbf{Linear 3} and the \textbf{Triangle}.

When breaking down the observed one-electron SIE into its components, we found that correlation usually played a minor role due to it being small and relatively constant. Exceptions were the SVWN5 and B97M-V approximations.
Most of the SIE stemmed from the approximate exchange functional and to a degree the one-electron error (OEE), which the is error in the expectation value of the one-electron Hamiltonian due to the incorrectly calculated density.
For example in \textbf{Linear 3} there was a substantial OEE, which then cancelled with a large portion of the exchange error, whereas the \textbf{Triangle} system had a larger exchange error and a much smaller OEE resulting in a larger total SIE.
Whenever two SIE components were involved in error compensation, it was usually between the exchange error and the OEE; this was also true for the larger geometries.

Through mean absolute error percentages (MAPEs) we showed that the majority of geometries were subject to functional errors with the exceptions of \textbf{Linear 3}, \textbf{Linear 5}, and the \textbf{Pentahedron}. Those cases were dominated by the density error. To further dissect this observation, we plotted  difference densities between a given DFA and HF to qualitatively visualize the SIE in each system. We discovered that in most systems the electron was equally shared over all nuclei for HF, and, that the SIE increased that delocalization \textit{equally} to all nuclei. Exceptions were again  \textbf{Linear 3}, \textbf{Linear 5}, and the \textbf{Pentahedron}, where the electron was semi-localized on one nucleus in the HF case with the DFT-based SIE shifting electron density from that one nucleus to the others, thus increasing the delocalization unequally. A possible conclusion is that the unequal electron sharing and the density error seem to be connected. As there has been limited information in the literature on what makes a system functional-error or density-error dependent, this unequal electron density delocalization may constitute a possible mode that causes the density error to appear in chemically more relevant systems. While we are not aware of any work that quantifies the density error and explicitly connects it to unequal electron sharing like we did, we found examples in the density-error literature that further support our statement. Larger density errors were described in systems for which we can infer unequal electron sharing, namely dissociations of NaCl, AlO, SiN$^+$, and CH$^-$,\cite{Kim2015} spin state differences in the [Fe(NCH)$_6$]$^{2+}$ complex,\cite{Burke_2018_2} or the barrier height in the oxygen transfer reaction between H and N$_2$O.\cite{Burke_2018_2}

In an alternative analysis, we showed through plotting the SIE versus the HF total energies that systems with more nuclei and higher symmetry seemed to suffer more from the SIE. This interpretation was supported by analyzing the crossover errors --- the difference in internuclear separation (distance) between where a density functional predicted the total energy of a given system to be zero compared to the exact HF solution within our chosen basis set. We showed that the SIE increased in magnitude going from smaller systems to larger ones.

Akin to our previous study,\cite{Lonsdale2020} we showed how higher nuclear charges induced a larger SIE in all systems.

Finally, we analyzed the effect of higher orbital occupation on mononuclear systems. 
We established that the SIE was greater in higher lying orbitals, which might be expected given the larger diffuseness of those higher lying states. This is particularly important in real-life applications where more diffuse valence orbitals are used to explain the underlying chemical effects.  Importantly, we also showed how seemingly one-electron SIE free methods in a 1s ground state are no longer SIE free for excited states. This is an important insight regarding the construction of DFAs even when they belong to the non-empirical class.

While our observed effects were exemplified for one-electron systems, they have relevance in treatments of chemically more relevant systems. This is why our findings are useful not only in the understanding of a long-standing problem in current DFT approaches, but also in trying to develop new approaches that are SIE free without sacrificing robustness in general applications.

\section*{Supplementary Material}
The Supplementary Material includes additional plots for all tested systems and functionals, including breakdowns of the SIE and MAPEs into their different components, the crossover-error analysis, SIE vs HF energy plots,  $Z$ dependence plots, and higher orbital dependence plots.

\section*{Data Availability Statement}
The data that supports the findings of this study are available within the article and its supplementary material. Raw data can be requested from the corresponding author.

\section*{Conflicts of interest}
The authors have no conflicts of interest to disclose.

\section*{Author Contributions}
Dale R. Lonsdale: conceptualization (equal); methodology (equal); investigation (lead); formal analysis (lead); writing - original draft (lead); writing - review and editing (equal).\\
Lars Goerigk: conceptualization (equal); supervision (lead); methodology (equal); resources (lead); writing - original draft (supporting); writing - review and editing (equal).

\section*{Acknowledgements}
Dale R. Lonsdale acknowledges an Australian Government Research Training Program Scholarship and is grateful to Sue Finch from the Statistical Consulting Centre for a helpful discussion about data analysis. Lars Goerigk is grateful for the generous allocation of resources by the National Computational Infrastructure (NCI) National Facility within the National Computational Merit Allocation Scheme (project ID: fk5).

\bibliographystyle{jcp}
\bibliography{SIE_Geoms_Orbitals}

\end{document}